\documentclass[fleqn,usenatbib]{mnras}

\usepackage{newtxtext,newtxmath}

\usepackage[T1]{fontenc}

\DeclareRobustCommand{\VAN}[3]{#2}
\let\VANthebibliography\thebibliography
\def\thebibliography{\DeclareRobustCommand{\VAN}[3]{##3}\VANthebibliography}

\usepackage{graphicx}	
\usepackage{amsmath}	
\usepackage{siunitx}
\usepackage{array}
\usepackage{pdflscape}
\usepackage{afterpage}
\newcommand{\mgii}{\ion{Mg}{II}}
\newcommand{\feii}{\ion{Fe}{II}}
\newcommand{\feiiems}{\ion{Fe}{II}$^{*}$}
\newcommand{\oii}{[\ion{O}{II}]}
\newcommand{\rcsa}{RCS0327}
\newcommand{\angstrom}{\mbox{\normalfont\AA}}
\newcommand{\MdotOut}{$\dot{M}_{out}$}
\newcommand{\MsolarPerYr}{$M_{\odot}\ yr^{-1}$}

\defcitealias{Shaban202230kpc}{Paper I}

\title[Dissecting a galactic outflow @ z $\sim$ 1.7]{Dissecting a 30 kpc galactic outflow at $z \sim $ 1.7}

\author[A. Shaban et al.]{Ahmed Shaban$^{1}$\thanks{E-mail: arshaban@ncsu.edu},
Rongmon Bordoloi$^{1}$\thanks{E-mail: rbordol@ncsu.edu},
John Chisholm$^{2}$,
Jane R. Rigby$^{3}$,
Soniya Sharma$^{4}$,
Keren Sharon$^{5}$,
\newauthor Nicolas Tejos$^{6}$,
Matthew B. Bayliss$^{7}$,
L. Felipe Barrientos$^{8}$,
Sebastian Lopez$^{9}$,
C\'{e}dric Ledoux$^{10}$, 
\newauthor Michael G. Gladders$^{11,12}$, 
Michael K. Florian$^{13}$
\\
$^{1}$Department of Physics, North Carolina State University, Raleigh, NC 27695-8202, USA\\
$^{2}$Department of Astronomy, University of Texas at Austin, Austin, TX 78712, USA\\
$^{3}$Observational Cosmology Lab, NASA Goddard Space Flight Center, Greenbelt, MD 20771, USA\\
$^{4}$Intel Corporation, 2501 NE Century Boulevard, Hillsboro, Oregon 97124, USA\\
$^{5}$Department of Astronomy, University of Michigan, 1085 South University Avenue, Ann Arbor, MI 48109, USA\\
$^{6}$Instituto de F\'isica, Pontificia Universidad Cat\'olica de Valpara\'iso, Casilla 4059, Valpara\'iso, Chile\\
$^{7}$Department of Physics, University of Cincinnati, Cincinnati, OH 45221, USA\\
$^{8}$Instituto de Astrof\'isica, Pontificia Universidad Catolica de Chile, Santiago, Chile\\
$^{9}$Departamento de Astronom\'ia, Universidad de Chile, Casilla 36-D, Santiago, Chile\\
$^{10}$ European Southern Observatory, Alonso de C\'ordova 3107,  Vitacura, Casilla 19001, Santiago, Chile\\
$^{11}$Department of Astronomy and Astrophysics, University of Chicago, 5640 South Ellis Avenue, Chicago, IL 60637, USA\\
$^{12}$Kavli Institute for Cosmological Physics at the University of Chicago, USA\\
$^{13}$Steward Observatory, University of Arizona, 933 North Cherry Ave., Tucson, AZ 85721, USA 
}

\date{Accepted 2023 September 29. Received 2023 September 25; in original form 2023 June 12}

\pubyear{2023}

\begin{document}
\label{firstpage}
\pagerange{\pageref{firstpage}--\pageref{lastpage}}
\maketitle

\begin{abstract}
We present the spatially resolved measurements of a cool galactic outflow in the gravitationally lensed galaxy {\rcsa} at $z \approx 1.703$ using VLT/MUSE IFU observations. We probe the cool outflowing gas, traced by blueshifted {\mgii} and {\feii} absorption lines, in 15 distinct regions of the same galaxy in its image-plane. Different physical regions, 5 to 7 kpc apart within the galaxy, drive the outflows at different velocities ($V_{out} \sim $ $-161$ to $-240$ km s$^{-1}$), and mass outflow rates ($\dot{M}_{out} \sim$ 183 to 527 {\MsolarPerYr}). The outflow velocities from different regions of the same galaxy vary by 80 km s$^{-1}$, which is comparable to the variation seen in a large sample of star-burst galaxies in the local Universe.  Using multiply lensed images of {\rcsa}, we probe the same star-forming region at different spatial scales (0.5 kpc$^2$ –- 25 kpc$^2$), we find that outflow velocities vary  between $ \sim $ $-120$ to $-242$ km s$^{-1}$, and  the mass outflow rates vary between $ \sim$ 37 to 254 {\MsolarPerYr}. The outflow momentum flux in this galaxy is $\geq$ 100\% of the momentum flux provided by star-formation in individual regions, and outflow energy flux is $\approx$ 10\% of the total energy flux provided by star-formation. These estimates suggest that the outflow in {\rcsa} is energy driven. This work shows the importance of small scale variations of outflow properties due to the variations of local stellar properties of the host galaxy in the context of galaxy evolution.

\end{abstract}




\section{Introduction}

Galactic outflows (or winds) are ubiquitous features in star-forming galaxies across cosmic time \citep{ veilleux2005galactic, rubin2010persistence, rubin2014evidence, Heckman2017GalacticWinds, rupke2018review} and are one of the cornerstones of the currently accepted galaxy evolution models \citep{somerville2015physical, Faucher-Giguere2023CGM}. Without these outflows, simulations fail to produce the current population of galaxies \citep{Dalla-Vcchia2012Simulating,Nelson2015FeedbackGasAccretion, Pillepich2018SimulatingGFormation}. Outflows regulate star-formation in galaxies \citep{Hopkins2010MaximumSFDensity} and, in some cases, quench the star-formation process \citep{hopkins2012stellar, hopkins2014galaxies, geach2018violent, man2018star, Trussler2020StarvationOutflowsQuench}. Outflows carry metals from the interstellar medium (ISM) of galaxies to the surrounding circumgalactic  medium (CGM)  \citep{Shen2013CGMGalaxyMassive,tumlinson2017circumgalactic, Peroux2020CosmicBaryon}, chemically enriching the CGM. This in-turn helps set the galaxy mass-metallicity relation \citep{tremonti2004origin, shen2012origin, Chisholm2018MetalEnrichedOutflow} and the stellar mass-halo mass relationship  \citep{Behroozi2013AverageSFH, Agertz2015SFandFeedbackSim}. The driving sources of these outflows can be stellar activities like: supernovae \citep{Lehnert1996Supernova-drivenWinds,Strickland2009SupernovaFeedbackM82}, and stellar winds from massive OB stars \citep{Heckman2017GalacticWinds}, or from active galactic nuclei (AGN) \citep{veilleux2005galactic,Ishibashi2022WhatPowersOutflows, Sorini2022BaryonsSIMBA}. In this paper, we focus on the outflows powered by star-formation \citep[e.g.,][for a review of the theory of star-formation driven outflows]{Zhang2018TheoryWindStellar}.

Galactic outflows are multi-phase, containing neutral gas, ions, molecular gas, and dust. These phases of baryonic matter exist at different temperatures covering a few orders of magnitude from $\sim\ \mathrm{few}\ - 100$ K (cold outflows), $10^4 - 10^5$ K (cool outflows), to $T > 10^6$ K (hot outflows) \citep{Somerville2015SFRwithMultiphaseGas, Thompson2016AnOriginMultiphase, Kim2020Framework4Multiphase, Fielding2022StructureMultiphase}. In order to observe all the relevant gas phases we need to target a plethora of wavelengths covering the electromagnetic spectrum from X-ray, UV, optical, to millimeter bands \citep[e.g.,][]{Sakamoto2014BipolarMolecularOutflows,bordoloi2014dependence, Zhu2015nearUV, Cazzoli2016NeutralOutlflowsOpticalNaD, chisholm2018feeding, Lopez2023XrayOutflowSFR}. All these phases can co-exist together in the same galaxy as multiphase outflow \citep[e.g., M82][]{Strickland2009SupernovaFeedbackM82}. They are also abundant in star forming galaxies \citep{rupke2005outflowsI, rupke2005outflowsII, rupke2018review}.

It is quite challenging to observe outflowing gas in the CGM directly through emission at high redshift, as the gas surface brightness is proportional to the square of the number density of the gas ($I \propto n^{2}$). This makes the emission too faint, and only the densest phases of the outflow will show emission. Studying the outflows in emission has been done in very bright objects \citep[e.g.,][]{rupke2019100, burchett2020circumgalactic, zabl2021muse, Shaban202230kpc, Dutta2023EmissionHalos}. Detecting the outflows in emission constrains the physical sizes and the geometries of these outflows, and how far they are from the galaxies generating them \citep{Wang2020OutflowsViaFluorescenceEmission}.

Another approach to study the faint outflows is through absorption lines. A background source, like a quasar or a bright galaxy, probes the foreground outflowing gas and the diffuse outflowing gas is detected as absorption against the light from the background source. This provides gas kinematics, column density, and line strength (equivalent width) of the outflowing gas along the line of sight \citep[e.g.,][]{rubin2010galaxies,bordoloi2011radial,bouche2012physical, schroetter2015VLTQSOProbeWinds, schroetter2016muse, Schroetter2019MEGAFLOWWindBackQuasar}. Owing to the paucity of bright background quasar or galaxy near a foreground galaxy, this method typically provides one sightline per galaxy, limiting these studies to characterizing the statistical properties in large samples of galaxies. 

A second method for studying the outflows in absorption is to use the stellar light of the galaxy itself to probe the outflows along the line of sight "Down the Barrel" \citep[e.g.,][]{adelberger2005connection,steidel2010structure,bordoloi2014modeling, rubin2014evidence, heckman2015systematic, chisholm2015scaling, chisholm2016shining, Xu2022CLASSYIII}. Both these methods by design only obtain global properties of outflows in these galaxies. In these studies, the kinematics and masses of galaxy wide outflows are correlated with the total star-formation rate (SFR) or the stellar mass ($M_{\star}$) of the galaxies to try to disentangle how they are driven. However, individual galaxies show large spatial variations of SFR and stellar-masses \citep{rigby2018magellanI, rigby2018magellanII}, suggesting that local star-forming properties might play an important role in driving star-formation driven outflows \citep{murray2005maximum}. Little information exists regarding spatial variation of galactic outflows within individual galaxies at these high-$z$ \citep{bordoloi2016spatially}.

There are three distinct approaches to study spatially resolved outflows in individual galaxies: (1) studying local galaxies with large angular footprint on the sky, e.g., Milky Way \citep{Fox2015FermiBubbles, Bordoloi2017FB, Clark2022GasFlowsMW} or M82 \citep{Strickland2009SupernovaFeedbackM82}, (2) studying other low redshift galaxies using spatially resolved IFU observations using rest-frame optical emission lines \citep{Sharp2010IFU10GWinds, ReichardtChu2022DuvetOutflows}, (3) to use gravitationally lensed galaxies which are naturally stretched over a large angular area on the sky. This allows us to study galactic outflow in a spatially resolved manner and trace the outflowing gas to their driving source \citep{bordoloi2016spatially,  james2018mapping, Sharma2018HighResLensed, fischer2019spatially, chen2021resolved}. Only gravitational lensing enables us to study spatially resolved properties of galaxies beyond the local universe.

Strong gravitational lensing works as a natural telescope for background galaxies. Lensing magnifies and stretches out the shape of distant galaxies while preserving their surface brightness. The apparent shape of galaxies in the image-plane (the plane on the sky in projection that contains the distorted and magnified images of the background galaxies) will appear larger than their true size in the source-plane (the plane where the background galaxy lies). Furthermore, the lensing can cause the background galaxy to have multiple images in the image-plane with varying magnifications. These multiple images magnify individual regions of the same galaxy at different physical sizes \citep{Schneider1992GravitationalLenses, Schneider2006gravitational}. We can use this effective ``zoom-in'' to trace the outflowing gas back to the local star-forming regions in a galaxy that drives them. By studying the relationship between spatially-resolved outflow properties in an individual galaxy and the local stellar properties of the star-forming regions that drive these outflows; one can create empirical models relating energy/momentum of the outflowing gas with the energy/momentum that drives these outflows \citep{heckman2015systematic}. This will inform the next generation of feedback models to create observationally realistic galaxies in numerical simulations.

The advent of wide field integral field unit (IFU) spectrographs has become a game-changer in efficiently carrying out spectroscopy of spatially extended faint gravitationally lensed arcs. IFU spectrographs can provide spectra for each location (spatial-pixel or spaxel) in the field of view of the unit. By observing the lensed galaxies with the IFUs, we can obtain spectral datacubes with good signal-to-noise ratios (SNR) in relatively short time compared to spectra from multiple single slits observations to cover the same field of view. From these IFU spectra, we can calculate the properties of the outflows and the properties of the local star forming regions and deduce if there are any correlations between them.

In this work, we study the spatial variation of the cool ($T \sim10^4$K) outflows in the strong gravitationally lensed galaxy RCSGA032727-132609 (hereafter referred to as {\rcsa}) at $z= 1.70347$  \citep{wuyts2010bright, wuyts2014magnified, Shaban202230kpc}. 
In \citet{Shaban202230kpc}, (hereafter \citetalias{Shaban202230kpc}), we measured the maximum spatial extent of the cool $10^4$ K outflowing gas to be $\approx$ 30 kpc using the resonant {\mgii} emission lines. In this work, we present the spatially resolved outflows kinematics and gas properties from blueshifted {\mgii} and {\feii} absorption lines.  The paper is organized as follows: \S \ref{sec:data_collect} describes the observations, \S \ref{sec:Methods} describes spectral extraction, and {\mgii} and {\feii} modeling, \S \ref{sec:results} describes the results of the analysis, \S \ref{sec:discuss} discusses the results in the context of the energetics of galactic outflows, and \S \ref{sec:conclusions} presents the conclusions. We assume a $\Lambda$CDM cosmology with a Hubble constant $H_0 = 70\ \rm{km\ s^{-1}Mpc^{-1} }$, matter density parameter $\Omega_m = 0.3$, and dark energy density parameter $\Omega_{\Lambda} =0.7$. We use the AB magnitude system in our calculations.

\begin{figure*}
    \centering
    \includegraphics[width=\textwidth]{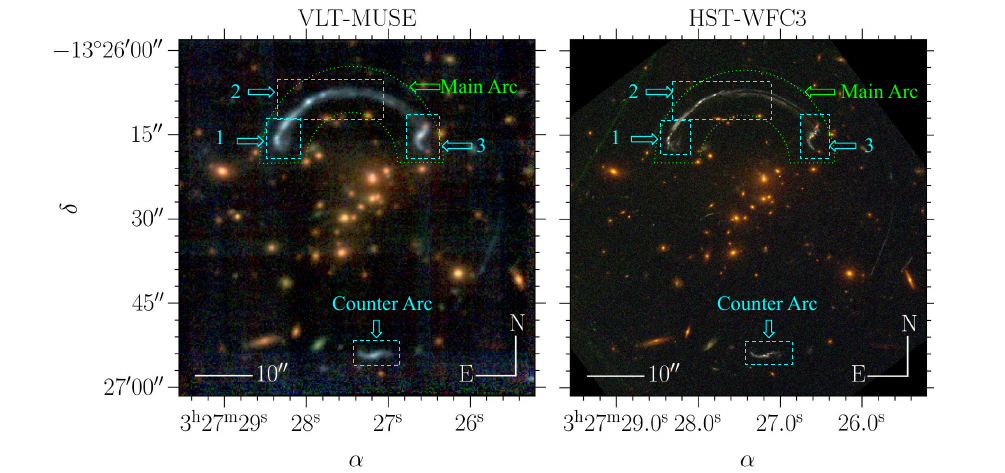}
    \caption{\textit{Left:} Colored image of {\rcsa} from the MUSE data cube. Three narrowband images of 100 {\AA} width are combined to create an RGB image. The blue, green and red narrow bands are centered at 4850 {\AA}, 6050 {\AA}, and 8050 {\AA}, respectively. \textit{Right:} Colored image of the same field using \textit{Hubble Space Telescope} (HST) Wide Field Camera 3 (WFC3) F390W, F606W, and F814W filters. The main arc is marked with a dotted green polygon. The dashed cyan rectangles mark the multiple images due to lensing of the target galaxy.}
    \label{fig:muse_hst}
\end{figure*}

\section{Description of Observations}\label{sec:data_collect}
{\rcsa} is a gravitationally lensed star-forming galaxy undergoing a merger at $z = 1.70347\pm 0.00002$ (see \S \ref{sec:GalaxyProperties}). The foreground galaxy cluster RCS2032727-132623 at $z = 0.564$ causes the strong gravitational lensing of the galaxy \citep{wuyts2010bright, gilbank2011red, wuyts2014magnified}. The apparent shape of the galaxy due to lensing consists of the main arc subtending  $38^{\prime\prime}$ on the sky \citep{wuyts2014magnified} and a smaller counter arc at the south of the lensing cluster (see Figure \ref{fig:muse_hst}). The galaxy is associated with a dark matter halo mass of $\mathrm{log_{10}(M_h/M_{\odot}) = 11.6\pm 0.3}$ \citep{bordoloi2016spatially}, corresponding to a circular velocity of $\mathrm{V_{cir}} = 156\pm 28\ \mathrm{km\ s^{-1}}$. We refer the reader to \citet{bordoloi2016spatially} for details of the arc. The left panel of Figure \ref{fig:muse_hst} shows the MUSE observation of {\rcsa}.

\subsection{MUSE Observations}\label{sub:MUSE_Observations}
 
We use observations of {\rcsa} with the Multi-Unit Spectroscopic Explorer \citep[MUSE;][]{bacon2010muse} on the Very Large Telescope (VLT) (program ID: 098.A-0459(A); PI: S. Lopez). {\rcsa} was observed for a total exposure time of 3.1 hours. We refer the reader to \citetalias{Shaban202230kpc} and \citet{lopez2018clumpy} for a detailed description of the observations and data reduction. The MUSE datacube is calibrated in air. We transform the wavelength of all the extracted spectra from the datacube from air to vacuum in the rest of our analysis.

\subsection{HST Observations}\label{sec:HST_Observations}

The Hubble Space Telescope (HST) Wide Field Camera 3 \citep[WFC3;][]{dressel2019wide} observations of {\rcsa} were used to construct the lens model \citep{sharon2012source}. These observations were obtained under HST/GO Program ID: 12267 (PI: J. Rigby) in the F606W, F390W, and F814W filters, respectively.  The exposure times in each filter are 1003s, 1401s, and 2133s for F606W, F390W, and F814W, respectively. The right panel of Figure \ref{fig:muse_hst} shows the RGB image representing {\rcsa} using these filters.   

We also use the HST-WFC3 photometric observations of the rest-frame $H\beta$ emission line and the continuum for {\rcsa} (Program ID: 12267, PI: J. Rigby, \citealt{wuyts2014magnified}) to calculate the star formation rates (SFR) and SFR surface densities ($\Sigma_{SFR}$; SFR per unit area). The $H\beta$ and the continuum were observed with the HST infrared filters F132N and F125W, respectively. The exposure times are 2112s and 862s for F132N and F125W, respectively. The specific details of these filters are summarized in \citet{dressel2019wide}.

\section{Methods}\label{sec:Methods}

\subsection{Extraction of 1D Spectra}\label{sec:spectra}
\begin{figure*}
    \centering    \includegraphics[width=\textwidth,height=\textheight,keepaspectratio]{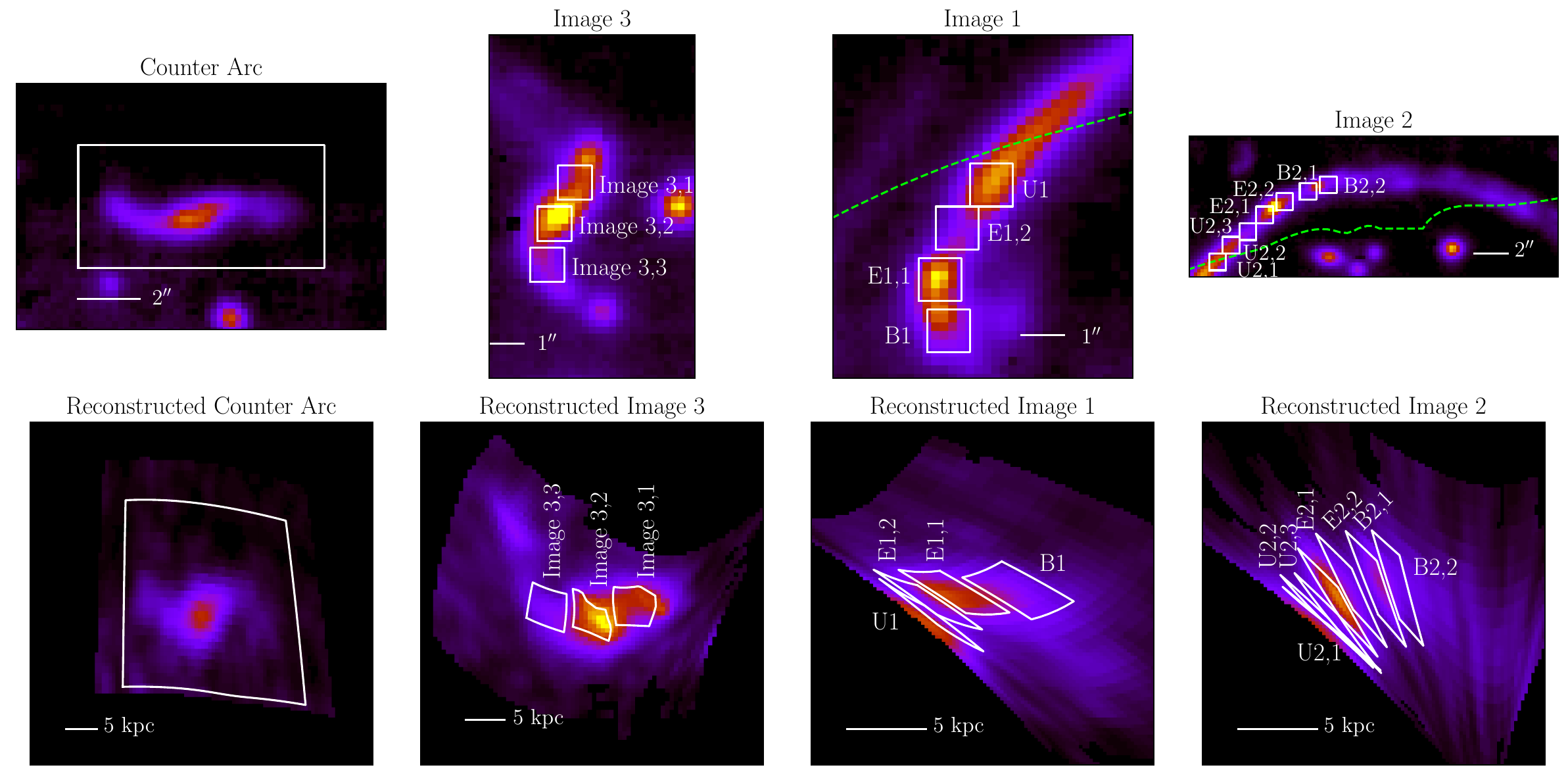}
    \caption{\textit{Top row:} MUSE white light images of the counter arc, image 3, image 1, and image 2 of {\rcsa} in the image-plane, respectively. The color map represents the surface brightness. Locations of pseudo-slits used for spectral extraction in the image-plane are shown as white polygons. The green lines represent the critical lines. The images are ordered in terms of the areas of the spectral extraction regions in the source-plane, with decreasing area and increasing magnification from left to right. \textit{Bottom row:} source-plane reconstructions of the images in the top row. The locations of the pseudo-slits in the source-plane are presented as white polygons. The labels in the panels represent the name of each pseudo-slit for the rest of the analysis.}
    \label{fig:Images123}
\end{figure*}

\begin{figure*}
    \centering
    \includegraphics[width=0.7\linewidth]{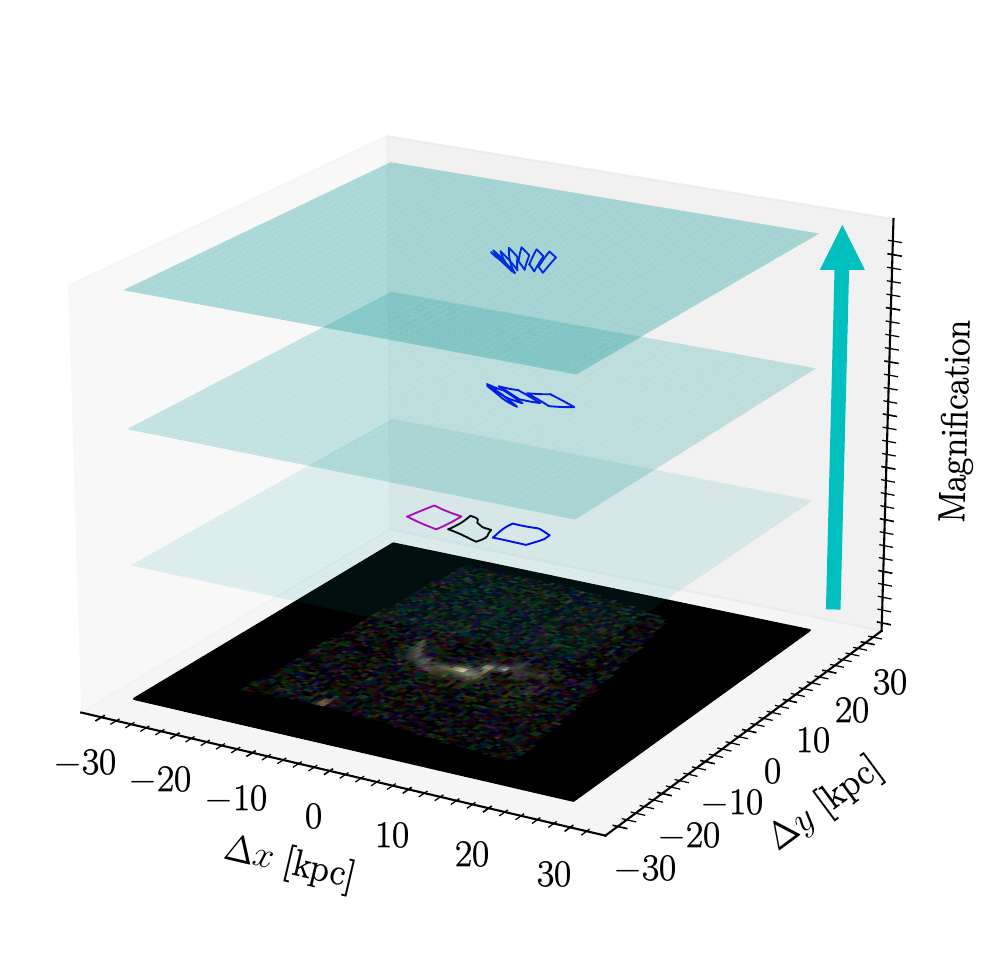}
    \caption{Location of the pseudo-slits in the source-plane of {\rcsa}. The x-y plane image represents the HST reconstruction of the counter arc image from Figure \ref{fig:muse_hst}. The z-axis represents the average lensing  magnification of each individual image in the image-plane, where magnification increases with higher $z$. The polygons represent the source-plane reconstructed pseudo-slits, similar to those in Figure \ref{fig:Images123}. The lower level (red, black, and blue polygons) is from image 3. The middle level (blue polygons) is from image 1, and the top level (blue polygons) is from image 2. Polygons of similar color are probing the same physical region with different magnifications. We use the lens model to do ray-tracing of the coordinates of the slits from the image-plane to the source-plane (see {\S} \ref{sec:GalaxyProperties}). The center of the x-y plane is the pixel with the highest surface brightness in the HST image in the image-plane.}
    \label{fig:3d_slits}
\end{figure*}

In this section, we describe how we extract and continuum-normalize the 1D spectra from the different regions of {\rcsa} from the MUSE datacube. First, we identify the highest surface brightness pixels along the curvature of the main arc. We then define 14 square pseudo-slits regions on the main arc, that are centered on these pixels. Each square is $1''\times1''$  ($5\times 5$ spatial pixels$^2$) in area and does not intersect with the neighboring squares. This makes all the selected pseudo-slits bigger than the maximum size of the seeing during the observation, which is $0.8^{\prime\prime}$. We choose these squares to be located on the star forming knots of the galaxy E, U, B, and G \citep{bordoloi2016spatially}. For the counter arc, we extract one spectrum that covers the whole arc. We are going to use this spectrum to represent the total integrated galaxy. These pseudo-slits  are shown in the image-plane of {\rcsa} in Figure \ref{fig:Images123}, top row. We ray-trace the position of these pseudo-slits onto the source-plane using the lens model (see \S \ref{sec:GalaxyProperties}) and present them in Figure \ref{fig:Images123}, bottom row. The location of these pseudo-slits relative to the galaxy in its source plane is presented in Figure \ref{fig:3d_slits}. The different x-y planes of Figure \ref{fig:3d_slits} correspond to the four images of the galaxy in the image-plane that have different magnifications and spatial distortions (see \S \ref{sec:discuss}).

\begin{figure*}
    \centering
    \includegraphics[width=\linewidth]{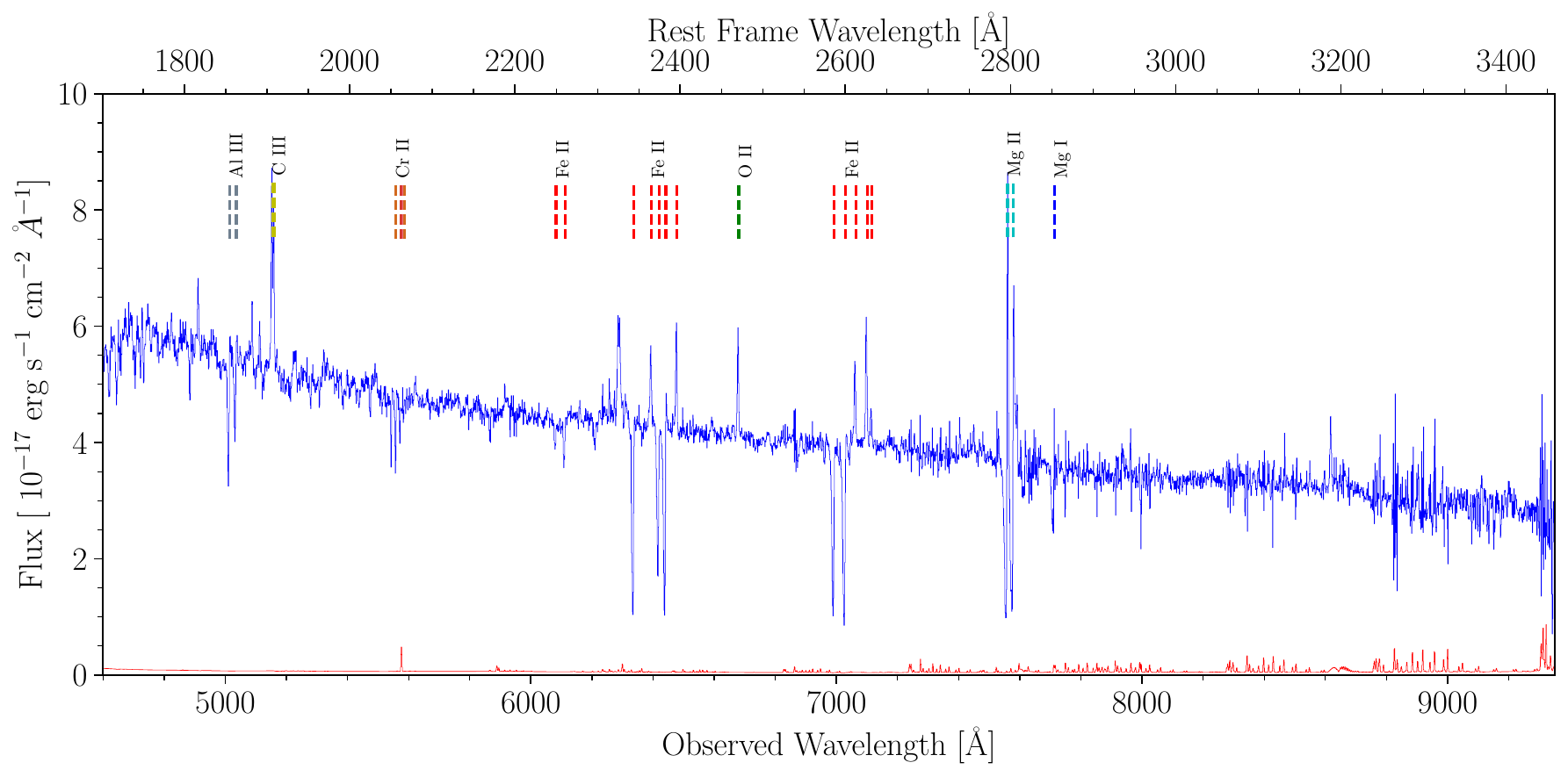}
    \caption{White light-weighted spectrum of the main arc from the MUSE data cube. The flux is shown as a solid blue line and the error is shown as a solid red line. The dashed colored lines represent the detected emission or absorption transitions in the spectrum, and each distinct color represents one ionic species. This integrated spectrum has a high SNR of 86, and even the weaker {\feii} 2249 line is detected. In our analysis, we exclusively use the {\oii}, {\feii}, and {\mgii} lines from MUSE observation. We mark the rest of the lines in the spectrum for completeness. All the transitions shown in this Figure are summarized in Table \ref{tab:lines_tot}.}
    \label{fig:lines_stack}
\end{figure*}

From each pseudo-slit, we create a 1D spectrum by summing the flux along the two spatial axes of the flux cube. The error spectrum is the square root of the sum of the variances from the variance cube for the same spatial regions. Then, we identify the important spectral lines in our spectra. The lines of interest in this work are {\mgii}, {\feii}, {\feiiems}, and {\oii}. The vertical dashed lines in Figure \ref{fig:lines_stack} annotate these lines, which are tabulated in more detail in Table \ref{tab:lines_tot}.

For each transition of interest, we select a $\pm$ 150 {\AA} region around it from the extracted 1D spectra.  We perform a local continuum fitting of these spectrum slices for each line by first masking all absorption and emission features, and then fitting a $\mathrm{3^{rd}}$ order polynomial to the continuum. We visually inspect each continuum fit and for each bad continuum fit we interactively perform a local continuum fit with a spline function using the \texttt{rbcodes}\footnote{https://github.com/rongmon/rbcodes} package \citep{rbcodes2022}.

We also produce a light-weighted spectrum of the main arc by weighting the spectrum of each individual spatial pixel by its surface brightness. Then, we sum these weighted spectra from the individual pixels to produce a light weighted spectrum for the main arc. We perform the same procedure for the counter arc too. The median SNR of the continuum around the transitions of interest is $\approx 86$ for the main arc, and $\approx 22$ for the counter arc. The weak {\feii} lines 2249.877, 2260.781 {\AA}, which are otherwise hard to detect, can be easily seen in the main arc and counter arc light-weighted spectra. Figure \ref{fig:lines_stack} shows the full light weighted spectrum of the main arc. We use a custom software made for this project, named \texttt{musetools} \footnote{https://github.com/rongmon/musetools}, to do the spectral extraction and the rest of data analysis.

\begin{table}
    \centering
    \begin{tabular}{cccc}
        \hline\hline
        Transition & $\lambda$ [{\AA}] & feature & $\mathrm{f_0}$\\
        \hline
        {\mgii}\ & 2796.351 & Resonant abs/ems & 0.6155 \\
         & 2803.528 & Resonant abs/ems & 0.3058 \\
        {\feii} & 2249.877 & Resonant abs & 0.001821 \\
         & 2260.781 & Resonant abs & 0.00244 \\
         & 2344.212 & Resonant abs & 0.1142 \\   
         & 2374.460 & Resonant abs & 0.0313 \\
         & 2382.764 & Resonant abs & 0.320 \\
         & 2586.650 & Resonant abs & 0.069125 \\
         & 2600.173 & Resonant abs & 0.2394 \\
       {\feiiems}\ & 2365.552 & Fine-structure ems & - \\
         & 2396.355 & Fine-structure ems & - \\
         & 2612.654 & Fine-structure ems & - \\
         & 2626.451 & Fine-structure ems & - \\
         & 2632.108 & Fine-structure ems & - \\
        \oii\ & 2470.97  & Emission & - \\
         & 2471.09  & Emission & - \\
         Al III & 1854.718 & Absorption & 0.559 \\
         & 1862.791 & Absorption & 0.278 \\
        C III] & 1906.683 & Emission & - \\
         & 1908.734 & Emission & - \\
        Cr II & 2056.257 & Absorption & 0.1030 \\
         & 2062.236 & Absorption & 0.0759 \\
         & 2066.164 & Absorption & 0.0512 \\
        Mg I & 2852.963 & Absorption & 1.83 \\
        \hline
    \end{tabular}
    \caption{Absorption and emission lines in the galaxy spectrum. $\lambda$ represents the vacuum wavelength of the transitions. $\rm{f_0}$ is the oscillator strength. These values are from \citet{morton2003atomic} and \citet{leitherer2011ultraviolet}.}
    \label{tab:lines_tot}
\end{table}

\subsection{Estimating Galaxy Properties}\label{sec:GalaxyProperties}

In this section we describe estimation of galaxy redshifts, star-formation rates and source-plane areas associated with the regions covered by each pseudo-slit. 

We need to quantify the systemic redshift for each individual region to set the zero systemic velocity of the galaxy, and to compute the relative Doppler shifts of absorption and emission lines in the analysis. We use the interstellar medium (ISM) fine-structure \oii\ emission doublet with vacuum wavelength $\lambda \lambda \ 2470.79, 2471.09$ {\AA} \citep{leitherer2011ultraviolet, tayal2007OIIFineStructure} to calculate the systemic redshift $z$ of each one of our chosen locations. 

We fit the doublet with a double Gaussian to the continuum normalized 1D spectra as shown below:
\begin{equation}
    F_{OII}(\lambda) = 1 + [G_1(\lambda,A,z,\Delta \lambda,\lambda_{1}) \ + G_2(\lambda,z,A_2,\Delta \lambda,\lambda_{2})],
\end{equation}
where $A$ is the flux amplitude of the first line, $A_2 (= c_{2,1} \times A)$ is the flux amplitude of the second line, $c_{2,1}$ is the line ratio between the second and the first line, \textit{z} is the systemic redshift, $\Delta \lambda$ is the line width of each line, $\lambda_1$ and $\lambda_2$ are the rest-frame wavelengths of the two lines ($\lambda_1 = 2470.97$ {\AA}, $\lambda_2 = 2471.09$ {\AA}). The line ratio $c_{2,1}$ must be in the range of $\approx$ 1 -- 2 according to their collision strength values at $T \sim 10^4 K $  \citep{tayal2007OIIFineStructure, kisielius2009OIIFineStructure}. The free parameters of this model are $A$, $c_{2,1}$, $\Delta \lambda$, and $z$. This model is convolved with a Gaussian with a full width at half maximum (FWHM) $\Delta \lambda = 2.7$ {\AA}, corresponding to the spectral resolution of the instrument at this wavelength. 

We use the Affine Invariant Markov chain Monte Carlo (MCMC) Ensemble sampler \texttt{emcee} \footnote{https://github.com/dfm/emcee} python package \citep{foreman2013emcee} to acquire the best constraints on the model parameters. \texttt{emcee} outputs the systemic redshift $z$ for each 1D spectrum as a posterior distribution. We take the $\rm 50^{th}$ percentile of this distribution (the median) as the systemic redshift, the $\rm 16^{th}$ and $\rm 84^{th}$ percentiles as lower and upper bounds for the redshift uncertainty, respectively. The best fit redshifts and their uncertainties are summarized in Table \ref{tab:galaxy_properties}.

To estimate SFR associated with each pseudo-slit, we use the $H\beta$ emission line from the HST narrowband imaging as a tracer of star formation. We calculate the zero point AB magnitude for the F132N to be 22.95 \citep{dressel2019wide}. We use this zero point for the rest of the work. We calculate the AB magnitude within each pseudo-slit by first summing the flux $f$ within them and accounting for the filter profile (see below). Then, we convert the flux to AB magnitude. 

For each chosen pseudo-slit, we choose a background region in the image-plane north of the arc without any light source to calculate the uncertainty on the AB magnitudes. This is repeated for all pseudo-slits. \citet{whitaker2014resolved} showed that for this galaxy, {\rcsa}, there is no significant variation in the extinction across the different star-forming regions using the $H\gamma / H\beta$ line ratio, and reported an average excess color value to be $E(B-V) = 0.4\pm 0.07$. We use this value to obtain the extinction-corrected AB magnitude in this analysis. 

We transform the extinction-corrected AB magnitude to flux density $f_{\nu}$ (units: $\rm{erg \ s^{-1} \ cm^{-2}\ Hz^{-1}}$). From the flux density and the frequency width of the filter, we obtain the flux in units of $\rm{erg\ s^{-1} cm^{-2}}$. The previous calculation gives flux values higher than the actual values because we are assuming that the HST-WFC3 are perfect square filters. To account for that, we calculate the flux of a synthetic Gaussian, with a given FWHM comparable to the FWHM of the {\oii} as a representative of the $H\beta$ emission line, using a square filter as done in the previous steps, and using the actual response throughput curve of the HST-WFC3 filter F132N from \texttt{pysynphot}\footnote{https://pysynphot.readthedocs.io/en/latest/} \citep{lim2015pysynphot}. We calculate the ratio of the flux from the square filter and actual response curve to be 2.15. Therefore, we divide the flux from the previous calculation by this ratio to obtain the actual flux values for each selected region.

These measured flux values are the image-plane values. To obtain the source-plane flux, we must account for the lensing magnification by the foreground cluster. We use the best available lens model, which was originally developed by \citet{sharon2012source} and later improved by \citet{lopez2018clumpy}. We refer the readers to these publications for a full description of the lens model. The lens model provides deflection and magnification maps, to relate the observed measurements to the un-lensed intrinsic properties of the source. We demagnify the image-plane extinction-corrected flux by dividing it by the median magnification at each pseudo-slit in the magnification map of the best-fit model from \citet{sharon2012source}. We also include the limits on magnification in each region, summarized in Table \ref{tab:galaxy_properties}, to account for the magnification uncertainty in the flux measurement. We use the extinction-corrected flux $F$ to calculate the $H\beta$ luminosity of each region as follows: $L(H_{\beta}) = 4 \pi D_{L}^{2} F$, where $D_{L}$ is the luminosity distance to the galaxy in the source-plane \citep{hogg1999distance}. To obtain the star formation rates from the luminosity, we assume that $L[H_{\alpha}] = 2.86\ L[H_{\beta}]$ \citep{osterbrock2006astrophysics}. We use the SFR equation from \citet{kennicutt1998star} for $H\alpha$. This can be expressed as:
\begin{multline}
    SFR (M_{\odot} yr^{-1}) = 7.9 \times 10^{-42} L[H_{\alpha}]\\ = 22.12 \times 10^{-42} L[H_{\beta}]
\end{multline}
This gives the star formation rate for each selected region. To estimate  the uncertainty on SFR, we propagate the uncertainties on AB magnitude, dust extinction, and magnification in all the steps described above. This provides us with the spatially resolved SFR measurements of the star-forming regions in {\rcsa} (see \citealt{whitaker2014resolved} for a similar analysis). The SFR values and the corresponding error bars are summarized in Table \ref{tab:galaxy_properties}.

\setlength{\tabcolsep}{10pt} 
\renewcommand{\arraystretch}{1.5}
\begin{table*}
    \centering
    \begin{tabular}{ccccccc}
        \hline
        \hline 
        Region & RA & Decl. & $z$ & Magnification & $SFR$ & $\Sigma_{SFR}$ \\
        Name &  &  &  &  & $[\rm{ M_{\odot}\ yr^{-1} }]$ & $[\rm{ M_{\odot}\ yr^{-1}\ kpc^{-2} }]$\\ 
        (1) & (2) & (3) & (4) & (5) & (6) & (7) \\
        \hline
        \hline
        B1 & 3:27:28.35 & -13:26:16.95 & $ 1.70318_{-0.00005}^{+0.00006}$ & $ 6.4_{-5.8}^{+7.1}$ &  $23_{-5}^{+6}$ & $2_{-2}^{+2}$ \\ 
        E1,1 & 3:27:28.36 & -13:26:15.75 & $ 1.70321_{-0.00002}^{+0.00002}$ & $ 10.2_{-8.9}^{+12.0}$ &  $94_{-22}^{+25}$ & $13_{-3}^{+4}$ \\ 
        E1,2 & 3:27:28.34 & -13:26:14.55 & $ 1.70338_{-0.00003}^{+0.00003}$ & $ 17.8_{-14.4}^{+22.9}$ &  $28_{-8}^{+9}$ & $7_{-2}^{+2}$ \\ 
        U1 & 3:27:28.28 & -13:26:13.55 & $ 1.70363_{-0.00002}^{+0.00002}$ & $ 33.2_{-23.7}^{+54.9}$ &  $23_{-10}^{+10}$ & $11_{-6}^{+6}$ \\ 
        \hline
        U2,1 & 3:27:28.21 & -13:26:12.55 & $ 1.70343_{-0.00002}^{+0.00002}$ & $ 120.5_{-57.5}^{+392.9}$ &  $4_{-5}^{+3}$ & $90_{-100}^{+66}$ \\ 
        U2,2 & 3:27:28.16 & -13:26:11.55 & $ 1.70367_{-0.00002}^{+0.00002}$ & $ 55.8_{-35.5}^{+129.5}$ &  $13_{-8}^{+8}$ & $9_{-6}^{+6}$ \\ 
        U2,3 & 3:27:28.09 & -13:26:10.75 & $ 1.70376_{-0.00003}^{+0.00004}$ & $ 30.6_{-23.1}^{+44.6}$ &  $19_{-7}^{+7}$ & $8_{-3}^{+3}$ \\ 
        E2,1 & 3:27:28.02 & -13:26:9.75 & $ 1.70334_{-0.00002}^{+0.00002}$ & $ 18.0_{-15.1}^{+22.5}$ &  $37_{-10}^{+11}$ & $9_{-2}^{+3}$ \\ 
        E2,2 & 3:27:27.94 & -13:26:8.95 & $ 1.70353_{-0.00003}^{+0.00003}$ & $ 14.0_{-12.2}^{+16.6}$ &  $56_{-13}^{+15}$ & $11_{-3}^{+4}$ \\ 
        B2,1 & 3:27:27.84 & -13:26:8.35 & $ 1.70296_{-0.00003}^{+0.00004}$ & $ 12.4_{-11.0}^{+14.4}$ &  $24_{-5}^{+6}$ & $4_{-1}^{+1}$ \\ 
        B2,2 & 3:27:27.76 & -13:26:7.95 & $ 1.70312_{-0.00003}^{+0.00004}$ & $ 11.7_{-10.4}^{+13.3}$ &  $29_{-6}^{+7}$ & $5_{-1}^{+1}$ \\ 
        \hline
        Image 3, 1 & 3:27:26.58 & -13:26:14.35 & $ 1.70355_{-0.00005}^{+0.00005}$ & $ 2.8_{-2.4}^{+3.5}$ &  $282_{-69}^{+85}$ & $14_{-4}^{+5}$ \\ 
        Image 3, 2 & 3:27:26.62 & -13:26:15.55 & $ 1.70371_{-0.00004}^{+0.00005}$ & $ 2.6_{-2.1}^{+4.2}$ &  $199_{-57}^{+89}$ & $11_{-3}^{+5}$ \\ 
        Image 3, 3 & 3:27:26.64 & -13:26:16.75 & $ 1.70519_{-0.00012}^{+0.00012}$ & $ 2.9_{-2.8}^{+3.1}$ &  $61_{-11}^{+14}$ & $3_{-2}^{+1}$ \\ 
        \hline
        Counter Arc & 3:27:27.18 & -13:26:54.35 & $1.70428_{-0.00005}^{+0.00006}$ & $2.6_{-0.1}^{+0.2}$ & $381_{-71}^{+86}$ & $6_{-1}^{+1}$ \\

        \hline
        \hline
    \end{tabular}
    \caption{Measured galaxy properties of the selected regions for spectral extraction pseudo-slits for RCS0327 as shown in Figure \ref{fig:Images123}. Column (1) is the location of each spectral extraction pseudo-slit with the name of the region (e.g. E, U, B, or Image 3), and the name of the image as shown in Figures \ref{fig:muse_hst} and \ref{fig:Images123}. Columns (2) and (3) are the right ascension and declination of the geometrical center of each polygon or pseudo-slit in Figure \ref{fig:Images123} in the image-plane. Column (4) is the measured redshift from the {\oii} nebular emission lines. Column (5) is the median magnification for each pseudo-slit as measured from the best lens model from \citet{sharon2012source, lopez2018clumpy}. Columns (6) and (7) represent the measured star formation rate (SFR) and SFR per unit area $\Sigma_{SFR}$ (or SFR surface density) in the source-plane. The lower and upper error bars represent the $\rm{16^{th}}$ and $\rm{84^{th}}$ percentiles limits on all the measured quantities. The horizontal lines in the Table represent the transition from one image to another image in the image-plane.}
    \label{tab:galaxy_properties}
\end{table*}

We divide the SFR in each pseudo-slit by its physical area in the source-plane to obtain the star-formation rate surface density $\Sigma_{SFR}$ (star-formation rate per unit area).

To calculate the area of each pseudo-slit in the source-plane, we use the lens model for this system as described above and ray-trace \citep{rbcodes2022,Bordoloi2022} the boundaries of the pseudo-slits using the lens equation:
\begin{equation}\label{eq:lens_equation}
    \vec{\beta}(\vec{\theta}) = \vec{\theta} - \vec{\alpha}(\vec{\theta})
\end{equation}
where $\vec{\theta}$ is the location in the image-plane, $\vec{\beta}$ is the corresponding location in the source-plane, and $\vec{\alpha}$ is the deflection for our source redshift in right ascension and declination derived from the best-fit lens model. For each location in the image-plane $\vec{\theta}$, there is a corresponding value for the deflection $\vec{\alpha}(\vec{\theta})$. We use the lens equation to produce Figure \ref{fig:3d_slits}. From the boundaries of each pseudo-slit in the source-plane, we can calculate the angular size covered by it. We then calculate the angular diameter distance $d_A$ to the source-plane. The angular size and the angular diameter distance are combined to estimate the physical area associated with each pseudo-slit in kpc$^2$.

The lens model has uncertainty on the deflection values $\vec{\alpha}(\vec{\theta})$. In order to propagate this deflection uncertainty to uncertainty on  $\Sigma_{SFR}$, we use 100 realizations of the lens model, including the best-fit model, and calculate the area for each realization. This gives us a distribution of values for the area. We take the 16$^\mathrm{th}$ and 84$^\mathrm{th}$ of that distribution as lower and upper bounds for the measured area in the source-plane. Then, it is straight forward to propagate the area uncertainty, and SFR uncertainty to the measured $\Sigma_{SFR}$. We show $\Sigma_{SFR}$ and the corresponding uncertainty in Table \ref{tab:galaxy_properties}.

\subsection{Modeling Outflows}\label{sec:modeling_outflows}
In this section, we describe the models used to characterize the {\mgii} and {\feii} lines to obtain the properties of the outflows. We model the absorption lines using two components, the systemic component $F_{sys}$ and the outflowing component $F_{out}$, respectively. The systemic component models the ISM contribution to the absorption features centered on the systemic zero velocity of the galaxy positions. The outflowing component models blueshifted outflowing gas with respect to the systemic zero velocity of the galaxy. We use a similar model as the one described in  \citet{ rupke2005outflowsI, sato2009aegis} and \citet{rubin2014evidence} for the systemic and outflowing components. For this model, we define the normalized flux as:

\begin{equation}
    F(\lambda) = 1 - C_{f}(\lambda) + C_{f}(\lambda) e^{-\tau(\lambda)},
\end{equation}
where $C_{f}(\lambda)$ is the gas covering fraction as a function of wavelength, and $\tau(\lambda)$ is the optical depth as a function of wavelength. We use a Gaussian to write the optical depth as:
\begin{equation}
    \tau(\lambda) = \tau_0 e^{-(\lambda - \lambda_0)^{2} / (\lambda_0 b_{D}/c)^{2}},
\end{equation}
where $\tau_0$ and $\lambda_0$ are the optical depth and rest-frame wavelength at the center of the Gaussian, respectively. $b_D$ is the Doppler velocity width of the absorption line, and $c$ is the speed of light in vacuum. For simplicity, we assume that $C_{f}(\lambda)$ for the outflowing component is constant, and does not vary with the wavelength (or velocity) for each region because of the low resolution of MUSE, and the low SNR for the optically thin absorption lines that produce robust estimates for $C_f(\lambda)$. For the systemic component, we assume that the absorbing gas fully covers the background starlight of our selected regions with $C_{f,sys}(\lambda) = 1$. We substitute $\tau_0$ using the relation between the optical depth $\tau_0$ and the column density $N$:
\begin{equation}
    N [cm^{-2}] = \frac{\tau_0 b_{D} [km \ s^{-1}]}{1.497\times 10^{-15} \lambda_{rest} [\angstrom] f_{0}}, 
\end{equation}
where $\lambda_{\mathrm{rest}}$ is the rest-frame wavelength, and $f_{0}$ is the oscillator strength of the absorption line \citep{Draine2011PhysicsISMIGM}. We do this step to implement the column density $N$ directly into our model to obtain a consistent value for the column density for all the absorption lines of the same species.
Since the outflowing component is by definition blueshifted, we multiply the optical depth of the outflow component with the function $\zeta$. We use this $\zeta$ to make the outflow component avoid including the redshifted absorption. It is defined as follows:
\begin{equation}\label{eq:eta}
    \zeta(\lambda, \lambda_{\mathrm{rest}}) = \left\{
                        \begin{array}{llll}
                            1 &; \lambda < \lambda_{\mathrm{rest}} &  & (\mathrm{blueshifted}) \\
                            0 &; \lambda \geq \lambda_{\mathrm{rest}} &  & (\mathrm{redshifted})
                        \end{array}
    \right.
\end{equation}
The free parameters for the outflow component are the central wavelength $\lambda_0$, the Doppler velocity width $b_{D}$, the covering fraction ${C_f}$, and the column density $N$, whereas for the systemic component $\lambda_0$ and ${C_f}$ (=1) are kept fixed.

Each emission line is modeled as a Gaussian as described below:
\begin{equation}
    F_{ems}(\lambda) = 1 + G(\lambda, A, v_0, b_D),
\end{equation}
where $v_0$ is the velocity at the center of each Gaussian, $A$ is the normalized flux amplitude at the center of each emission line, and $b_D$ is the Doppler velocity width of the emission component. The normalized flux amplitudes of the  emission line of the same ionic species will be tied to the first line in wavelength order in the component through the line ratios ($c_{ij}=\frac{A_i}{A_j}$, where $i$ and $j$ represent two emission lines of the same ionic species). The free parameters for the emission are the central velocity $v_0$, Doppler velocity width $b_D$, the flux amplitudes $A$, and the line ratios $c_{ij}$.

The full model that fits the spectrum in each region is the multiplication of all the three components,
\begin{equation}
    F(\lambda) = \left( F_{out}(\lambda) \times F_{sys}(\lambda) \times F_{ems}(\lambda) \right) \ast G_{LSF}(\lambda),
\end{equation}
which is then convolved with a Gaussian $G_{LSF}(\lambda)$ that has FWHM corresponding to the spectral resolution of MUSE. 

For {\feii} lines in the observed frame $\lambda_{obs} = 6220 - 6490$ {\AA}, the spectral resolution is $\Delta \lambda \approx$ {2.63 \AA} (or $\Delta v \approx 123 \ \mathrm{km \ s^{-1}}$), and for the {\feii} lines in the observed frame $\lambda \sim 7030$ {\AA }, the spectral resolution is $\Delta \lambda \approx$ {2.56 \AA} (or $\Delta v \approx 110 \ \mathrm{km \ s^{-1}}$). For the {\mgii} lines in the observed frame $\lambda_{obs} \approx 7560$ {\AA}, the spectral resolution is $\Delta \lambda \approx$ {2.52 \AA} (or $\Delta v \approx 100 \  \mathrm{km \ s^{-1}}$).

\begin{figure*}
    \includegraphics[width=\textwidth,height=\textheight,keepaspectratio]{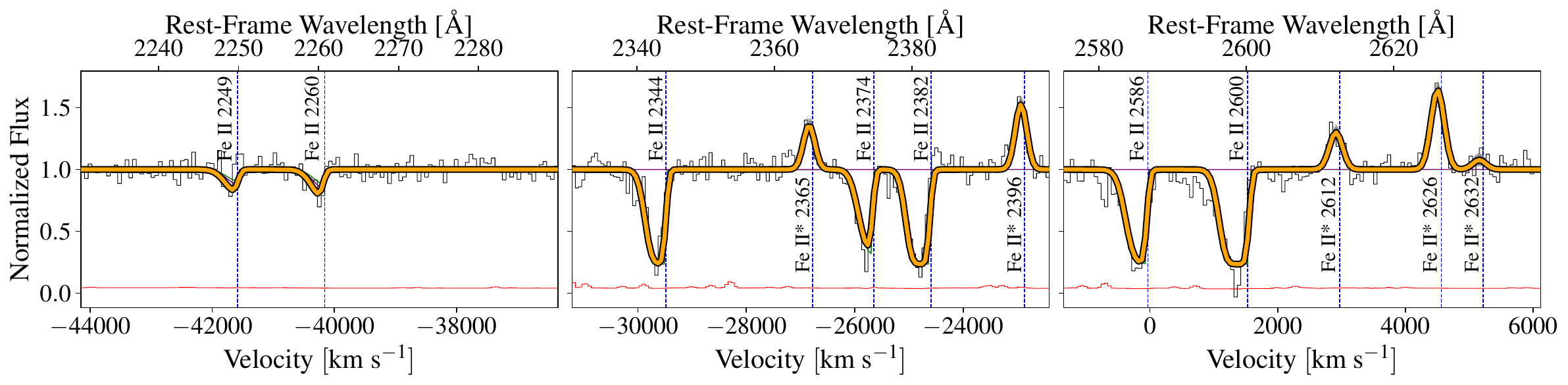}
    \caption{Simultaneous fit of {\feii} absorption and {\feiiems} emission lines to the extracted spectrum of the counter arc. Best fits to the {\feii} lines around the rest-frame wavelength $\lambda \approx 2250$ {\AA} (left panel),  $\lambda \sim$ 2300-2400 {\AA} (middle panel) and  $\lambda \approx$ 2600 {\AA} (right panel) are presented. The black and red solid lines represent the continuum-normalized flux and error, respectively. The orange line represents the final best fit model described in Section \ref{sec:modeling_outflows}. The gray thin lines show 300 random realizations of the \texttt{emcee} runs of the model parameters. The {\feii} 2600 {\AA} line's blueshifted velocity wing is contaminated by the Mn II 2594.499 {\AA} line. Therefore, we do not use this {\feii} line for the outflow velocity estimate in our analysis. }
    \label{fig:FeII_fitting}
\end{figure*}

\begin{figure}
    \centering
    \includegraphics[width=\linewidth]{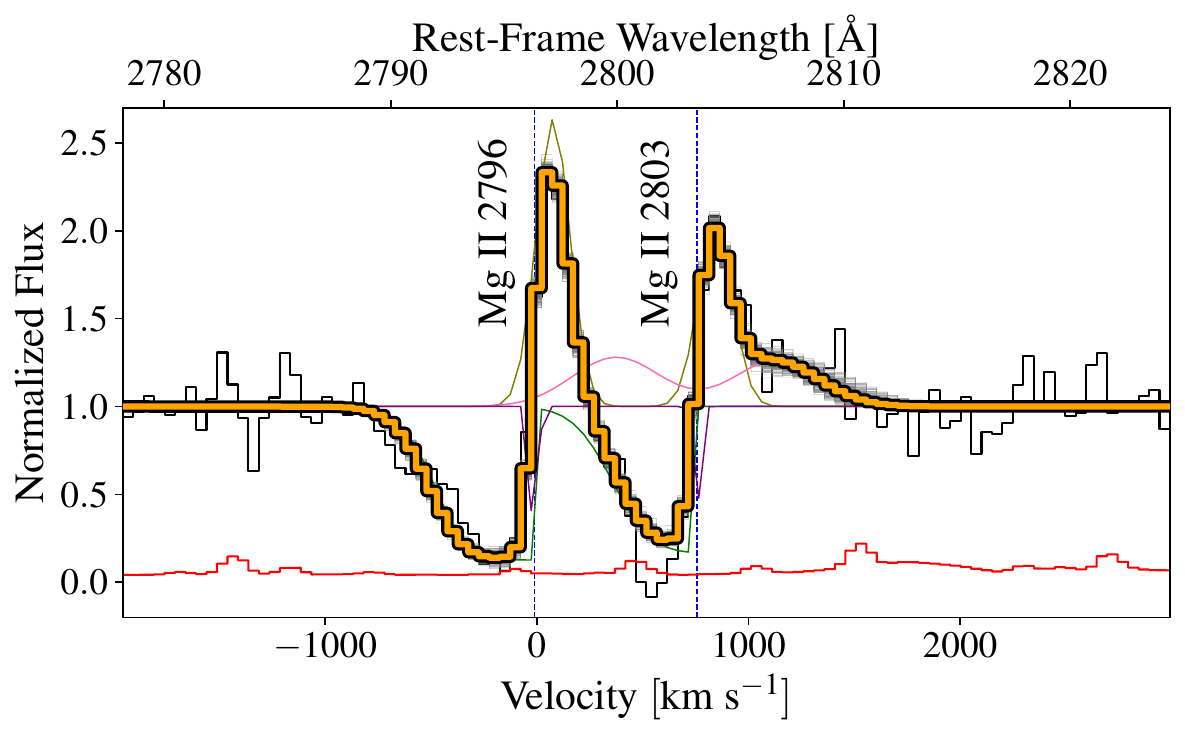}
    \caption{Simultaneous fit of {\mgii} absorption and emission lines to  the counter arc spectrum. The black and red lines represent the continuum-normalized flux and error spectrum, respectively. The orange line is the best fit model. The gray thin lines show 300 random realizations of the \texttt{emcee} runs.The thin green and purple lines are unconvolved ouflowing and systemic components, respectively. The olive and pink lines show the primary and secondary emission components, respectively. The {\mgii} 2803 {\AA} absorption line in the counter arc has a high uncertainty on the flux.}
    \label{fig:MgII_fitting}
\end{figure}

We constrain the best fit model parameters by sampling the posterior probability density function (PPDF) for each model using the \texttt{emcee} package. The number of walkers used is 50, and the total number of steps is 30000 for {\feii} modeling and 50000 for {\mgii}. We discard about 10\% of the steps of each chain as a burn-in time. We apply uniform priors to allowed parameter intervals that are adjusted for each individual model. The best-fit model for both {\feii} and {\mgii} absorption and emission lines in the counter arc spectrum are shown in Figures \ref{fig:FeII_fitting} and \ref{fig:MgII_fitting}, respectively. 

We calculate the acceptance fraction as described in \citet{foreman2013emcee} for each model to make sure that the MCMC chains are converging. All the measured acceptance fractions are within the recommended range of 0.2 and 0.5 \citep{foreman2013emcee}. 

For each realization, we compute the rest-frame equivalent width $W_r$ to quantify the line strength of the outflowing and emission components, given by:
\begin{equation}\label{EW_def}
    W_r = \int (1 - F(\lambda) )d\lambda, 
\end{equation}
where $F(\lambda)$ is the normalized flux as a function of rest-frame wavelength. This gives us a distribution in values of $W_r$. We calculate the $\mathrm{50^{th}}$ percentile (the median) as the best fit equivalent width for the line, and the $\mathrm{16^{th}}$ and $\mathrm{84^{th}}$ percentiles as the lower and the upper bounds of $W_r$, respectively.

Similarly, for each realization, we calculate the absorption weighted outflow velocity for each transition as:
\begin{equation}
    V_{out} = \int v (1 - F_{out}(v)) dv,
\end{equation}
where $F_{out}(v)$ is the normalized flux as a function of velocity for the outflowing component. We take the $\mathrm{50^{th}}$ percentile of the $\mathrm{V_{out}}$ distribution as the mean absorption weighted outflow velocity for that transition. The $\mathrm{16^{th}}$ and $\mathrm{84^{th}}$ percentiles of the $\mathrm{V_{out}}$ distribution are taken as the lower and upper bounds of $\mathrm{V_{out}}$, respectively.

For the outflows traced by the {\feii}, we use the transition {\feii} 2382 {\AA} as the indicator of the outflow properties $W_r$ and $\mathrm{V_{out}}$ as it has the highest oscillator strength $f_0$. Therefore, this transition shows optically thin high velocity absorption wings that can not be detected using weaker {\feii} transitions. For the same reason, we use the $\lambda$ = 2796.351 {\AA} transition for outflow properties traced by {\mgii} absorption. As the high velocity blue-shifted absorption of {\mgii} $\lambda$ {2803.528 \AA} gets emission filled by the resonant {\mgii} 2796 {\AA} emission feature (Figure \ref{fig:MgII_fitting}), the $\lambda = $ 2796.351 {\AA} gives the cleanest measurement of outflowing gas traced by {\mgii} absorption. The {\mgii} emission shows a secondary peak at $\rm{\sim 385_{-36}^{+46}\ km\ s^{-1} }$ with respect to the systemic velocity in Figure \ref{fig:MgII_fitting} with a broad Doppler parameter $\rm{b_D = 296_{-56}^{+60}\ km\ s^{-1}}$. This secondary peak shows up clearly in 11 spectra of the individual pseudo-slits of the main arc. The other 3 spectra correspond to parts of regions E and B, probed by image 1 and image 2 in the image-plane.

\begin{figure*}
    \centering
    \includegraphics[width=\linewidth]{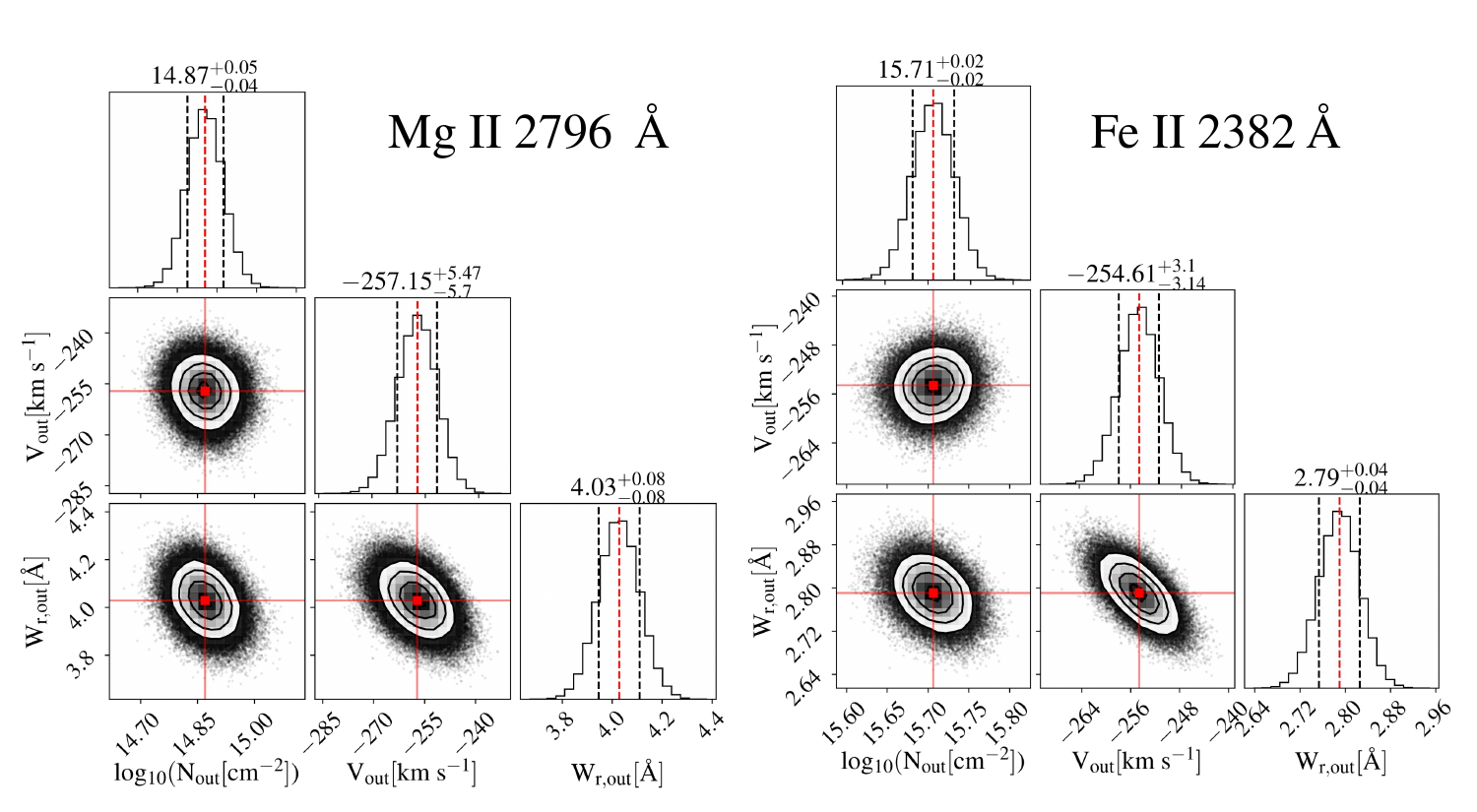}
    \caption{Posterior probability distribution functions (PPDFs) showing the outflow properties for both {\mgii} (\textit{Left}) and {\feii} (\textit{Right}) from the counter arc spectrum. PPDFs of outflow velocity ($\mathrm{V_{out}}$), outflow column density ($\mathrm{\log_{10}( N_{out})}$) and outflow equivalent widths ($\mathrm{W_{r,out}}$) are presented. These measurements represent the average global outflow properties in the galaxy. The outflow velocity measurement from {\mgii} and {\feii} are consistent with each other within the error bars. The error-bars on $\rm{log_{10}(N_{out})}$ represent the uncertainty from model fitting. The {\mgii} column densities are poorly constrained due to saturation of the lines (see Figure \ref{fig:MgII_fitting}).}
    \label{fig:corner_measured_outflows}
\end{figure*}

Figure \ref{fig:FeII_fitting} and Figure \ref{fig:MgII_fitting} show the best fit model profiles of the {\feii} and {\mgii} lines obtained by fitting the light weighted 1D spectrum of the counter arc, respectively. The gray lines show the 300 random realizations of the model created by drawing from the PPDF. The {\feii} absorption lines, show seven transitions {\feii} 2249.877, 2260.781, 2344.212, 2374.460, 2382.764, 2586.650, and 2600.173 {\AA}, out of which two are optically thin ({\feii} 2249, 2260 {\AA}). Five fluorescent {\feiiems} emission lines are also detected: {\feiiems} 2365.552, 2396.355, 2612.654, 2626.451, and 2632.108 {\AA} (see Table \ref{tab:lines_tot}). We simultaneously fit all these lines, and the presence of optically thin absorption lines allow robust estimates of {\feii} outflow column densities. 
Figure \ref{fig:corner_measured_outflows} (right panel) shows the PPDF of the estimated {\feii} outflow column density, outflow velocity, and outflow rest-frame equivalent width for the counter arc. We measure the median outflow velocity to be $\rm{-254\pm 3\ km\ s^{-1}}$, the outflow rest-frame equivalent width to be $\rm{W_{r, out} = 2.79\pm 0.04}$ {\AA}, and the outflow column density to be $\rm{ log_{10}(N [cm^{-2}]) = 15.71\pm 0.02}$. The covering fraction is one of the model parameters for the outflow component. The inferred $C_{f,out}$ for {\feii} in the counter arc spectrum is $0.77\pm 0.01$.

The strong {\mgii} absorption lines are saturated as measured by the rest-frame equivalent width ratios. Both absorption lines show similar equivalent widths $W_{r, 2796} = 4.03\pm 0.08$ {\AA} and $W_{r, 2803} = 3.05\pm 0.07$ {\AA} corresponding to line ratio of $1.32\pm 0.02$ for the counter arc spectrum. This combined with the emission filling arising from strong resonant {\mgii} emission in Figure \ref{fig:MgII_fitting} results in more uncertain estimates of gas column densities in the {\mgii} transitions. The left panel of Figure \ref{fig:corner_measured_outflows} shows the PPDF of the estimated {\mgii} outflow column density, outflow velocity, and outflow equivalent width. The median outflow velocity is $\rm{-257_{-6}^{+5}\ km\ s^{-1}}$, the rest-frame equivalent width is $\rm{W_{r,out} = 4.03\pm 0.08}$ {\AA}, and the outflow column density is $\rm{log_{10}(N_{out}[cm^2]) = 14.87_{-0.04}^{+0.05} }$. The error bars on $\rm{log_{10}(N_{out})}$ for {\mgii} here represent the uncertainty from the model fitting process and not the actual uncertainty due to saturation of the absorption lines, which is expected to be larger than the model fit uncertainty. It is worth noting that the median outflow measurements from both {\feii} and {\mgii} absorption lines in the counter arc spectrum are consistent with each other within error bars. The inferred covering fraction for the outflow component $C_{f,out}$ for {\mgii} is $0.88\pm 0.02$. The full measurements of outflow properties around individual regions of {\rcsa} are presented in Table \ref{table:outflow_properties}.

\section{Results}\label{sec:results}
In this section, we describe the variation of outflow kinematics and strengths traced by both emission and absorption lines across the arc {\rcsa}. We ray-trace the location of each spaxel of the pseudo-slits from the image-plane to the source-plane as described in Section 3.2 and measure the physical separation between them. Throughout the rest of the paper, we always discuss the variation of outflow properties in the source-plane of the galaxy.

\subsection{Outflow Absorption Strength and Kinematics}\label{sec:out_results}
\begin{figure*}
    \centering
    \includegraphics[width=0.8\linewidth]{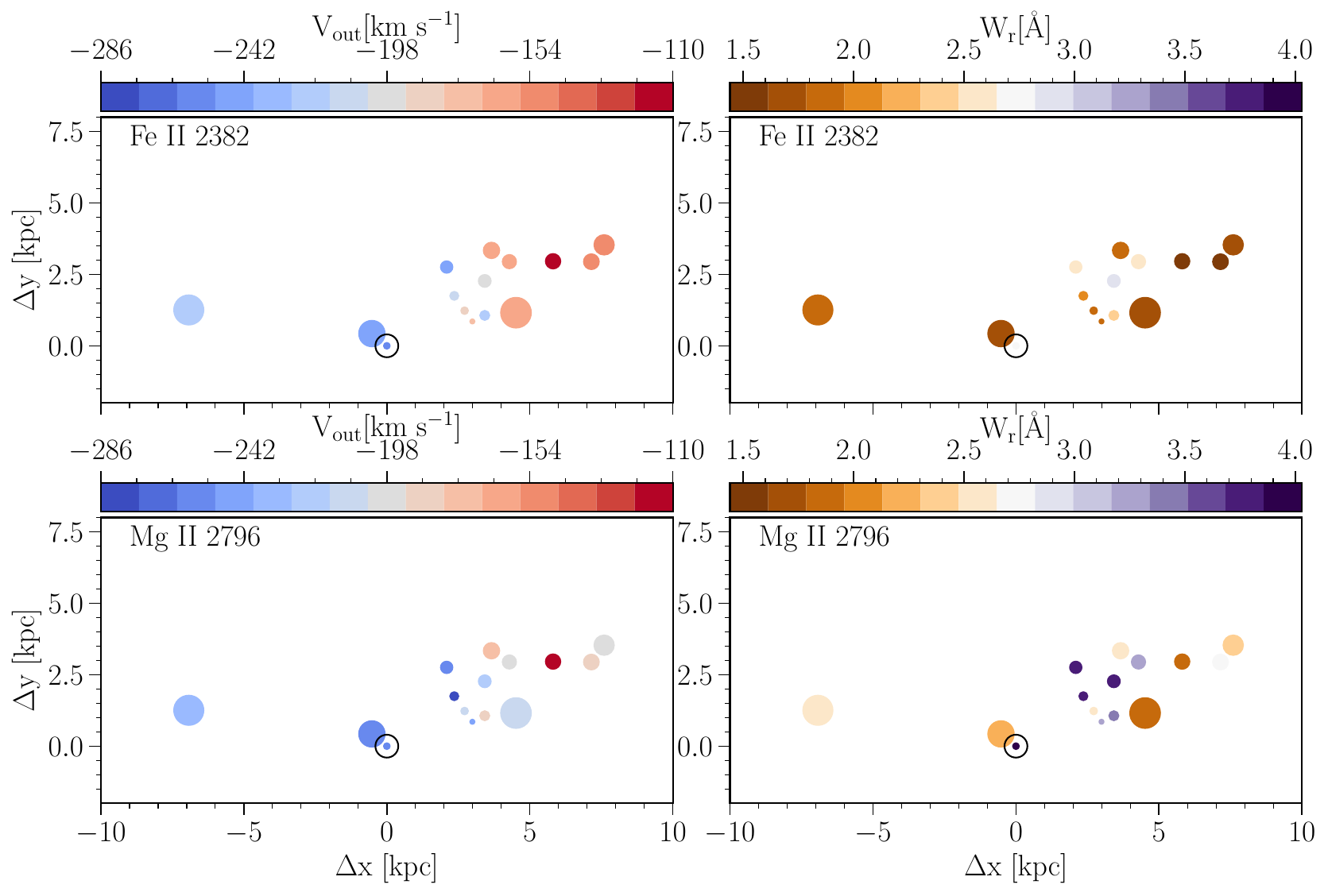}
    \caption{Outflow properties and the locations of individual regions in the source-plane. The x- and y-axes represent the physical distance in the source-plane for all panels. \textit{Top row:} Outflows properties traced by {\feii} 2382. \textit{Bottom row:} Outflows properties traced by {\mgii} 2796. \textit{Left:} The median outflow velocity. \textit{Right:} Rest-frame equivalent width for the outflow component. The sizes of the data points represent the areas of each selected region in the source-plane. The data point with a black open circle around it in each subplot represents the global value from the spectrum of the counter arc. }
    \label{fig:MgII_FeII_source}
\end{figure*}

We first present the spatial variation of outflow kinematics and absorption strengths across different parts of the galaxy {\rcsa}. Figure \ref{fig:MgII_FeII_source} shows the variation of mean outflow velocities $\mathrm{V_{out}}$ (left columns) and rest-frame equivalent widths $W_r$ (right columns) of {\feii} and {\mgii} absorption, in the source-plane of {\rcsa}. The circles show their location in the source-plane of the galaxy (relative to the galaxy center) and are color-coded to show  $\mathrm{V_{out}}$ (left panels) and $W_r$ (right panels), respectively. The open circle show the measurement for the ``global" outflow properties obtained from the light-weighted spectrum of the counter arc. The right (west) side of the galaxy is probed at different physical scales by the pseudo-slits multiple times, which enables us to ``zoom-in" at the same physical region of the galaxy and study the local variation of outflowing gas at different spatial location of the galaxy (see Figure \ref{fig:3d_slits}).

Across the arc, the {\feii} outflow velocities vary from $-120 \mathrm{km\ s^{-1}}$ to more than $-242 \mathrm{km\ s^{-1}}$, and the {\mgii} outflow velocities vary from $-110\ \mathrm{km\ s^{-1}}$ to $-287\ \mathrm{km\ s^{-1}}$, which is a factor of two change across $\sim 15$ kpc of the galaxy. The mean outflow velocity of all the individual pseudo-slits across the main arc is $-185\ \rm{km\ s^{-1}}$ and the standard deviation is 35 $\rm{km\ s^{-1}}$ for the {\feii}. For the {\mgii}, the mean outflow velocity across the main arc is $-212\ \rm{km\ s^{-1}}$, with a standard deviation of 42 $\rm{km\ s^{-1}}$.

The integrated outflow velocity towards the counter arc (see Figure \ref{fig:muse_hst}) is $\mathrm{- 254\ km\ s^{-1}}$ for {\feii}, and $\mathrm{- 257\ km\ s^{-1}}$ for {\mgii}, respectively. These values are $\rm{70\ km\ s^{-1}}$ larger than the measured mean outflow velocity of all the individual regions of the main arc. This counter arc measurement is akin to a typical  ``down-the-barrel" outflow experiment  (e.g., \citealt{chisholm2016shining, Xu2022CLASSYIII}), where one obtains the 1D spectrum of a galaxy by integrating all the light coming from the galaxy in the slit. It is worth noting that, for this individual galaxy, depending on where one is measuring the outflow velocity, the inferred velocity might vary by a factor of two. Without gravitational lensing (aided by highly efficient IFUs), performing spatially resolved long-slit measurements on different parts of faint high-$z$ galaxies become prohibitively expensive.

Figure \ref{fig:MgII_FeII_source} (right column) shows the variation of {\feii} 2382 (top row) and {\mgii} 2796 (bottom row) outflow absorption strengths across {\rcsa}. Both the {\feii} and {\mgii} absorption strengths (rest-frame equivalent width $\rm{W_r}$) vary by a factor of $\approx 2$ across 15 kpc of the galaxy. {\feii} outflow absorption strengths are  1.4 {\AA} $\mathrm{ < W_r< 2.9}$ {\AA}, and for {\mgii} outflow absorption strengths are 1.8 {\AA} $\mathrm{< W_r< 3.8}$ {\AA}, respectively. The measured outflow absorption strengths and velocities are summarized in Table \ref{table:outflow_properties}.

The median outflow velocities of {\feii} and {\mgii} that trace the bulk of the outflowing gas (Figure \ref{fig:MgII_FeII_source}) are below the escape velocity ($\mathrm{V_{esc} = 418_{-89}^{+112}\ km\ s^{-1}}$; \citealt{bordoloi2016spatially, Klypin2011DarkMatterHalos}; see Appendix \ref{appendix:escape_velocity}) at 5 kpc. This 5 kpc distance is comparable to the measured scale distances from \citetalias{Shaban202230kpc} (see Table \ref{tab:mean_radius_outflow} in Appendix \ref{appendix:Outflow_Properties}). This suggests that if the bulk of the cool outflowing gas is within 5 kpc, it will be retained by the gravitational potential well of the galaxy.

However, the extreme blueshifted wings of the absorption lines of the outflowing gas might have different behaviour. To quantify that, we calculate the $\mathrm{90^{th}}$, and $\mathrm{95^{th}}$ percentile outflows velocities and check if at these extreme velocities some part of the outflowing gas is escaping the galaxy's gravitational potential. We measure these velocities from the cumulative distribution functions (CDFs) of the best fit outflow models. These values are summarized in Table \ref{table:outflow_properties} in Appendix \ref{appendix:Outflow_Properties}. Figure \ref{fig:v50_90_95_MgII_FeII} shows these velocities for {\feii} in the top row, and for the {\mgii} in the bottom row. 

For the {\feii} transition, the $\mathrm{90^{th}}$ percentile velocities between individual pseudo-slits ranges from $-212\ \rm{km\ s^{-1}}$ to $-453\ \rm{km\ s^{-1}}$ with an average value of $-327\ \rm{km\ s^{-1}}$, and standard deviation of 67 $\rm{km\ s^{-1}}$, respectively. The global value (from the counter arc) for the $\mathrm{90^{th}}$ percentile outflow velocity for {\feii} is $-390\ \rm{km\ s^{-1}}$. This is higher than the average of the individual pseudo-slits probing different regions of the galaxy. Similarly, the $\mathrm{95^{th}}$ percentile outflow velocities from the individual pseudo-slits ranges from $-236\ \rm{km\ s^{-1}}$ to $-501\ \rm{km\ s^{-1}}$ with an average value of $-362\ \rm{km\ s^{-1}}$, and standard deviation of 76 $\rm{km\ s^{-1}}$, respectively. The global value for the $\mathrm{95^{th}}$ percentile outflow velocity is $-434\ \rm{km\ s^{-1}}$, which is again higher than the average of the individual pseudo-slits and also higher than the escape velocity (> $418\ \rm{km\ s^{-1}}$) at 5 kpc.

For the {\mgii} transition, the $\mathrm{90^{th}}$ percentile outflow velocity in different pseudo-slits of the main arc ranges from $-208\ \rm{km\ s^{-1}}$ to $-531\ \rm{km\ s^{-1}}$ with an average value of $-386\ \rm{km\ s^{-1}}$ and standard deviation of 82 $\rm{km\ s^{-1}}$, respectively. The measured global $\mathrm{90^{th}}$ percentile outflow velocity from the counter arc is $-496\ \rm{km \ s^{-1}}$, which is again higher than the average of the individual measurements. Unlike the {\feii} $\mathrm{90^{th}}$ percentile outflow velocity from the counter arc, the $\mathrm{90^{th}}$ percentile velocities traced by {\mgii} are higher than the escape velocity even at 5 kpc. For the more extreme $\mathrm{95^{th}}$ percentile velocity, the scatter from the individual pseudo-slits from the main arc ranges from $-231\ \rm{km\ s^{-1}}$ to $-608\ \rm{km\ s^{-1}}$ with an average of $-431\ \rm{km\ s^{-1}}$ and standard deviation of 96 $\rm{km\ s^{-1}}$. The global value for the $\mathrm{95^{th}}$ outflow velocity as measured from the counter arc is $-563\ \rm{km\ s^{-1}}$, which is $\approx$ 130 $\rm{km\ s^{-1}}$ faster than the average of individual regions of the main arc. 

These findings suggest that the high velocity outflow components (traced by $\mathrm{90^{th}}$ or $\mathrm{95^{th}}$ percentile outflow velocities) are consistent with escaping the gravitational potential well of {\rcsa}. It should be stressed that these absorption components represent only a small fraction of the total outflowing absorption, and the bulk of the outflowing gas is consistent with being bound to the dark matter halo of the host galaxy. Furthermore, these findings highlight that there exists significant variability between the outflow velocity from location to location.

From Figure \ref{fig:v50_90_95_MgII_FeII}, we notice that many of the selected regions' $\mathrm{90^{th}}$ and $\mathrm{95^{th}}$ percentile velocities are large enough to exceed the escape velocity at 5 kpc. This small portion of the cool outflow will escape the galaxy potential well and will end up at much larger radii in the CGM, or may end up enriching the intergalactic medium (IGM) with metals.

The covering fraction $C_{f,out}$ of the outflowing component also shows variation across the main arc. The {\feii} lines have more robust estimates of $C_{f,out}$ due to the detection of the optically thin lines. For the individual pseudo-slits with area less than 12 kpc in images 1 and 2, $C_{f,out}$ has a range of 0.63--0.95, a mean value of 0.74, and standard deviation of 0.1 as measured from {\feii} lines. The corresponding pseudo-slit that overlaps with all of them and with an area of $\approx$ 25 kpc has $C_{f, out}$ of 0.67 from {\feii} lines. We notice that the mean $C_{f,out}$ of the individual small regions is comparable to the counter arc value of $0.77\pm 0.01$. 
Similarly, for the {\mgii} lines in the individual pseudo-slits in images 1 and 2, $C_{f,out}$ is in the range of 0.77--0.96. The mean $C_{f,out}$ for these individual pseudo-slits is $0.88$, and the standard deviation is 0.6 for {\mgii}. The larger pseudo-slit from image 3 with an area 25 kpc that and overlaps with these pseudo-slits has a $C_{f, out}$ value of 0.53, which is 0.33 offset from the mean value of the individual pseudo-slits. The mean $C_{f,out}$ of the individual is almost identical to $C_{f,out}$ from the counter arc, with a value of $0.88\pm 0.02$. The values for the covering fraction $C_{f,out}$ for {\feii} and {\mgii} are available in Table \ref{table:outflow_properties} in Appendix \ref{appendix:Outflow_Properties}.
This shows that the outflows covering the SF regions in the galaxy have different sizes and covering fractions of the slits, depending on the area and the region probed in the galaxy. This will have further impact on the measured properties of the outflow, like the mass outflow rate $\dot{M}_{out}$.

\begin{figure*}
    \centering
    \includegraphics[width=0.8\textwidth]{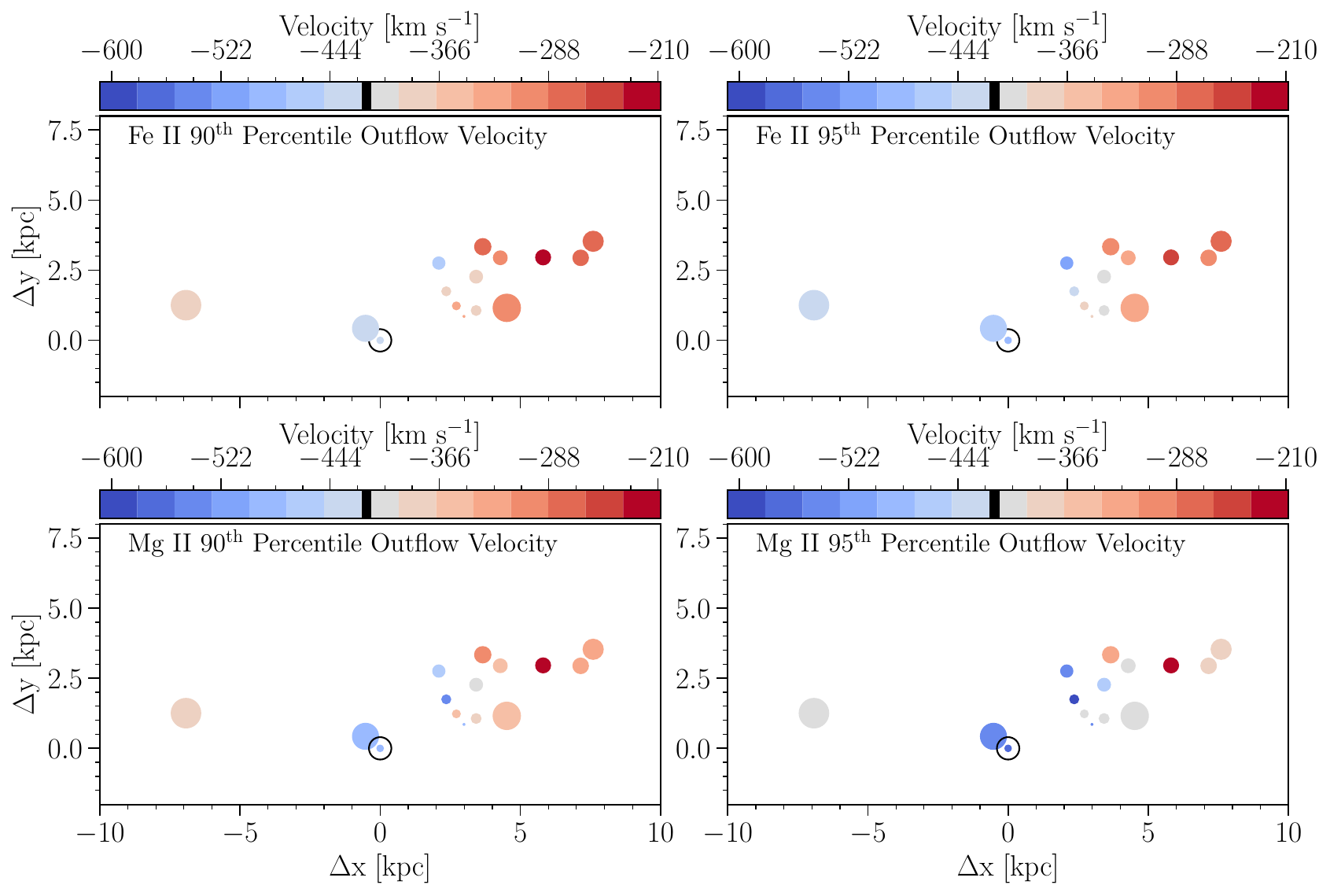}
    \caption{Extreme outflow velocities in the source-plane. The x- and y-axes represent the physical distance in the source-plane for all panels. \textit{Top row:} The $\rm{90^{th}}$ (\textit{Left}), and the $\rm{95^{th}}$ (\textit{Right}) percentile velocities for the {\feii} outflow component. \textit{Bottom row:} The $\rm{90^{th}}$ (\textit{Left}), and the $\rm{95^{th}}$ (\textit{Right}) percentile velocities for the {\mgii} outflow component. The colors of the data points represent the value of the velocity (see color bar above plots). The sizes of the data points represent the areas of the selected regions in the source-plane. The data points with black open circles around them represent the global values from the spectrum of the counter arc. The black vertical band in each color-bar represents the value for the escape velocity ($\rm{ V_{esc} = -418\ km\ s^{-1}}$) at 5 kpc. } 
    \label{fig:v50_90_95_MgII_FeII}
\end{figure*}

\subsection{{\mgii} and {\feii} emission line properties}

In each individual 1D spectrum of {\rcsa}, five fluorescent {\feiiems} emission lines, and two resonant {\mgii} emission lines are detected as summarized in Table \ref{tab:lines_tot}. These emission lines are spatially extended and in particular, diffuse {\mgii} emission is detected out to 30 kpc of {\rcsa}. We refer the reader to \citetalias{Shaban202230kpc}, where the detailed properties of this spatially extended emission are described. 

The {\feiiems} lines originate owing to the de-excitation of the {\feii} atoms, where the electron moves from the excited level to one of the levels resulting from the fine-structure splitting of the ground state. These fine-structure levels have slightly higher energy compared to the original ground state level \citep{prochaska2011simple}. This makes the {\feiiems} fluorescent emission lines appear in the spectrum at redder rest-frame wavelengths, compared to the {\feii} resonance absorption lines. We enforce the emission velocity of all the {\feiiems} emission lines to be the same when performing the model fits. The {\feiiems} measurements  summarizing emission velocity $\mathrm{V_{ems}}$, Doppler parameter $\mathrm{b_D}$, and the rest-frame equivalent width $\mathrm{W_r}$ (of  {\feiiems} 2626) are presented in Table \ref{tab:outflows_emission}.

\begin{figure*}
    \centering
    \includegraphics[width=\linewidth]{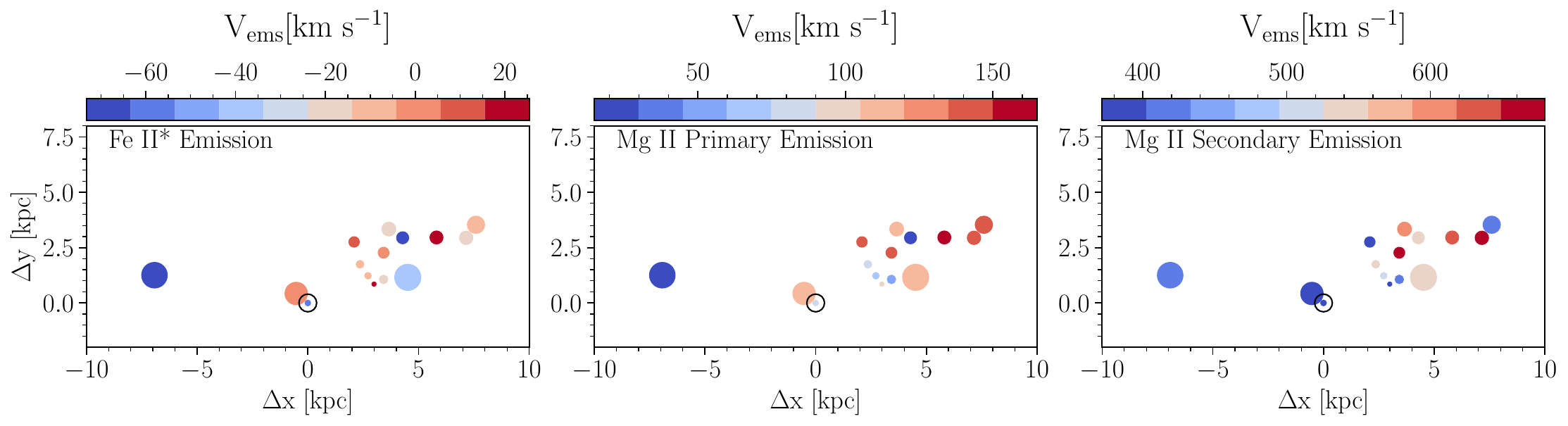}
    \caption{The emission velocities in the source-plane. \textit{Left:} Emission velocity as traced by {\feiiems} emission line 2626 {\AA}. \textit{Middle:} Velocity of the first emission component of {\mgii} back-scattered emission. \textit{Right:} Velocity of the second emission component of {\mgii} back-scattered emission. The data points with black open circles around them represent the global values from the spectrum of the counter arc. The sizes of the data points are proportional to their areas in the source-plane. The color-bars scales in the 3 panels are different to show the variation in each one.}
    \label{fig:emission_vel_MgII_FeII}
\end{figure*}

Figure \ref{fig:emission_vel_MgII_FeII}, left panel, shows the measured velocities of the emission lines at the locations of the pseudo-slits in the source-plane, similar to Figure \ref{fig:MgII_FeII_source}. The emission velocity measurements for {\feiiems} are scattered around the zero velocity of the galaxy. The {\feiiems}'s $\mathrm{V_{ems}}$ takes values from $\mathrm{-73\ km\ s^{-1}}$ to $\mathrm{25\ km\ s^{-1}}$ for the main arc pseudo-slits, and takes a value of $\mathrm{-59_{-3}^{+3}\ km\ s^{-1}}$ for the counter arc. The redshifted (positive) $\mathrm{V_{ems}}$ are from 3 pseudo-slits representing parts of regions E, U, and B in images 1 and 2, which are star-forming local regions in the galaxy. However, $\mathrm{V_{ems}}$ for image 3 and the counter arc are blueshifted. Even within the same galaxy, the {\feiiems} emission line kinematics can differ by $\rm{\approx 100\ km\ s^{-1}}$.

For the {\mgii} resonant emission, two distinct redshifted emission components are seen in the {\mgii} 2803 transition. We model them using two emission components that are interpreted as scattered emission from the back side of the outflowing gas. This emission traces the densest regions of the outflows. We interpret these emission components as two distinct redshifted shells of outflows. These distinct shells of outflowing gas may have originated from two distinct episodes of star-burst that took place in the past. The middle and right panels of Figure \ref{fig:emission_vel_MgII_FeII} show the emission velocities for the {\mgii} primary component and the secondary component, respectively.

The velocity of the first {\mgii} emission component is redshifted and varies from $\mathrm{15\ km\ s^{-1}}$ in image 3 to  $\mathrm{165\ km\ s^{-1}}$ in region B of image 2. The global value for velocity for the primary emission component in the counter arc is $\mathrm{87\ km \ s^{-1}}$. For the secondary {\mgii} emission component, the velocities have much higher redshifted velocity values compared to the primary component, ranging from $\mathrm{372\ km \ s^{-1}}$ for region B in image 2 to  $\mathrm{679\ km \ s^{-1}}$ for region E in image 2. The global value for the velocity of this component is $\mathrm{ 385\ km\ s^{-1}}$ as measured from the counter arc. All these measurements show that there is an average velocity difference between the two components $\approx 400\ \mathrm{km\ s^{-1}}$. This is consistent with the detection of these two components in the surface brightness emission maps in \citetalias{Shaban202230kpc}.

\subsection{Mass Outflow Rates}\label{sec:MassOutRates}
For outflowing gas with total hydrogen volume density $n(R)$, mean molecular weight $\mu m_p$,  outflow velocity $v(R)$, and spatial extent $R$, the mass outflow rate is:

\begin{equation}
    \begin{array}{ll}
      \dot{M}_{out} & = \int_{A} \mu m_{p}\ \ n(R) \ v(R) dA \\ & = \int_{A} \mu m_{p}\ \ n(R) \ v(R) \ R^2 \ d\Omega,   
    \end{array}
\end{equation}
where $\Omega$ is the solid angle covered by the outflow. Assuming a thin shell geometry, $n$ and $v$ are constant within each shell at that distance. Then the expression reduces to:

\begin{equation}
    \dot{M}_{out} = \Omega_w \ \mu  m_p\ n \ R^2 \ v= \Omega_w \ \mu m_p\ N_H \ R \ v,
\end{equation}
where $N_H$ is the total column density along the line of the sight.

The mass outflow rates are highly uncertain and depend on accurate measurement of  spatial extent $R$, and $N_H$. As discussed in \S \ref{sec:modeling_outflows}, we measure outflow column densities of {\mgii} and {\feii} transitions, respectively. As our observations cover many optically thin transitions of {\feii} (see Table \ref{tab:lines_tot}), the {\feii} column densities are robust. Whereas all {\mgii} transitions are optically thick, and their column densities are formally lower limits.

Using these measurements, we estimate a conservative lower limit on the total hydrogen column density of the outflowing gas by assuming no ionization correction ($N_{MgII} = N_{Mg}$, and $N_{FeII} = N_{Fe}$). We assume that the cool outflowing gas metallicity to be the ISM metallicity  \citep{wuyts2014magnified}, as the cool gas is most likely ejected with material from the ISM in the star forming regions. It could also be mixed with metal-enriched gas from SNe (as suggested in \citealt{Chisholm2018MetalEnrichedOutflow}). For dust depletion, a depletion factor of -1.3 dex for Mg and a factor of -1.0 dex for Fe is adopted from \citet{jenkins2009unified}. The solar abundances values for {\mgii} and {\feii} are adopted from \citet{lodders2003solar} and \citet{Asplund2009SunComposition}. Using these considerations, we compute a lower limit of $N_H$ for each pseudo-slit.

The last uncertainty in estimating mass outflow rate comes from the uncertainty in knowing the spatial extent of the outflow ($R$). In \citetalias{Shaban202230kpc}, we show that the {\mgii} emission profile around {\rcsa} is spatially extended, and report a radial surface brightness profile. Since, this {\mgii} emission originates in the back scattered light of the galactic wind, it represents the spatial extent of the outflow (of at least the densest part of the outflowing gas). The mean light weighted spatial extent of the outflowing gas ($\langle R \rangle $) is estimated as 
\begin{equation}
   \langle R \rangle =\int_{R_0}^{R_f} \ R \ P(R) \   dR,
\end{equation}
where $R_0$ is assumed to be the radius of the star forming regions ($\approx 100$ pc), and $R_f$ is the radius that marks the end of the measured surface brightness radial profile, and $P(R)$ is the normalized {\feiiems} or {\mgii} surface brightness profile adopted from \citetalias{Shaban202230kpc}. The typical values for $\langle R \rangle$ for {\feiiems} and {\mgii} are $\geq 4.7$ kpc. These values summarized in Table \ref{tab:mean_radius_outflow} in Appendix \ref{appendix:Outflow_Properties}. This changes the expression for minimum mass outflow rates to

\begin{equation}
    \dot{M}_{out} \gtrsim 22 M_{\odot} \ yr^{-1} \frac{\Omega_w}{4\pi} \ \frac{v}{300\ \rm{km\ s^{-1}}}  \frac{N_H}{10^{20}\rm{cm^{-2}}}\ \frac{\langle R \rangle}{5\ \rm{kpc}}. 
\end{equation}

\begin{figure*}
    \centering
    \includegraphics[width=\textwidth]{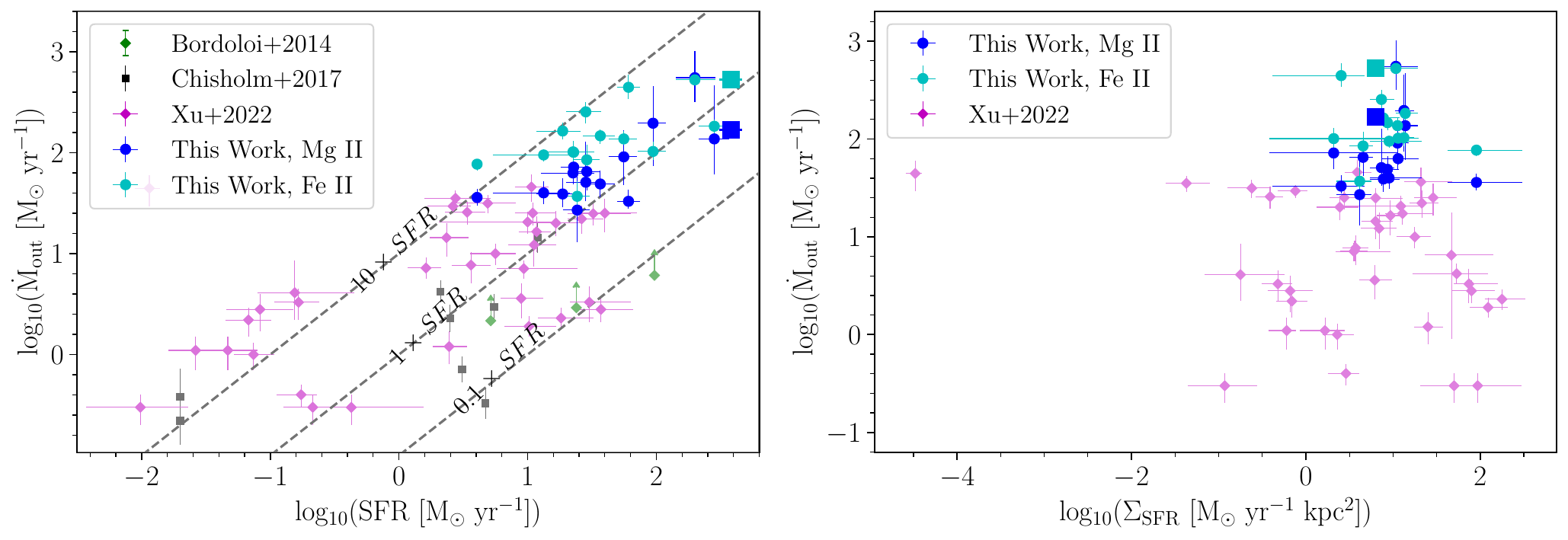}
    \caption{\textit{Left}: Mass outflow rate {\MdotOut} versus the SFR for {\mgii} (blue circles), and {\feii} (cyan circles) for {\rcsa} in log scale. The blue and cyan squares in both panels represent the measurements for the counter arc for {\mgii} and {\feii}, respectively. The magenta diamonds and gray squares are from \citet{Xu2022CLASSYIII} and \citet{chisholm2017GalacticOutflows}, respectively. The green diamonds are lower limits for $\dot{M}_{out}$ from \citet{bordoloi2014dependence}. The black diagonal lines in the panel represent the mass loading factor values ($\eta = \dot{M}_{out}/SFR$ = 0.1, 1, 10). \textit{Right}: Mass outflow rates versus SFR per unit area $\Sigma_{SFR}$, in log scale. The colors of the data points are the same as the left panel. In both panels, we see the consistency with respect to the trends of the data reported in \citet{Xu2022CLASSYIII}. Since the $\dot{M}_{out}$ measurements from {\rcsa} are lower limits, this implies that {\rcsa} is likely located along the mass loading factor of $\eta \sim 10$ as seen from the individual regions. }
    \label{fig:Mdot_out_SFR_SSFR}
\end{figure*}

Figure \ref{fig:Mdot_out_SFR_SSFR}, left panel, shows the measured mass outflow rates traced by {\mgii} (blue circles) and {\feii} (cyan circles) transitions versus SFR for individual pseudo-slits. The three diagonal black lines in the left panel correspond to the mass loading factor ($\eta \; = \dot{M}_{out}/SFR$) with values of 0.1, 1, and 10, respectively. We see a scatter in the mass outflow rates, with the highest {\MdotOut} of more than $527_{-47}^{+48}$ {\MsolarPerYr} and the lowest {\MdotOut} of $37_{-11}^{+15}$ {\MsolarPerYr} in the {\feii} transition. The global value for {\MdotOut} is $\approx$ $169_{-17}^{+20}$ {\MsolarPerYr} for {\mgii} (blue square), and $527_{-29}^{+30}$ {\MsolarPerYr} for {\feii} (cyan square) as measured from the counter arc spectrum. We compare these measurements with a sample of star-forming galaxies in the local universe from \citet{Xu2022CLASSYIII} (magenta diamonds; $z < 0.2$), \citet{chisholm2017GalacticOutflows} (gray squares; $z < 0.2$), and at higher redshift from \citet{bordoloi2014dependence} (green diamonds; $1 \leq z \leq 1.5$). It is clear that the {\MdotOut} measured in individual local regions of {\rcsa}, are significantly higher than that of the previous studies' data points. The scatter in {\MdotOut} measured in one galaxy {\rcsa} is similar to what is observed for the full sample of local star-burst galaxies and stacks of galaxies at $1 \leq z \leq 1.5$. From the diagonal dashed lines in the left panel in Figure \ref{fig:Mdot_out_SFR_SSFR}, we also note that the majority of the sightlines towards {\rcsa} have mass loading factors between $\sim$ 1--10. By contrast, the  mass loading factor for the counter arc is $\approx 1$. This shows that measurement of {\MdotOut} by taking integrated flux of the whole galaxy might underestimate the true {\MdotOut} around individual regions of a galaxy, which might be an order of magnitude higher. All the outflow rates values are summarized in Table \ref{table:M_P_E_dot} in Appendix \ref{appendix:Outflow_Properties}.

Figure \ref{fig:Mdot_out_SFR_SSFR}, right panel, presents the mass outflow rates ({\MdotOut}) versus star-formation rate per unit area ($\Sigma_{SFR}$). We also see a scatter in the relation between {\MdotOut} and $\Sigma_{SFR}$ similar to what is seen from samples of star-forming galaxies from \citealt{Xu2022CLASSYIII} ($z \leq 0.2$). The dynamic range of $\Sigma_{SFR}$ probed for {\rcsa} is not large enough to see any definite trend of {\MdotOut} and $\Sigma_{SFR}$ for our measurements. These large differences in {\MdotOut} measurements within a single galaxy highlight the importance of understanding the small-scale local properties of the galaxies, which drive these galactic outflows.

\section{Discussion}\label{sec:discuss}

\subsection{Top-Down Look:}

An advantage of studying a gravitationally lensed galaxy is that different parts of the same galaxy are multiply imaged at different magnifications. Depending on their location relative to the caustic, some images capture the whole scale of the galaxy (e.g., image 3 and the counter arc image in Figure \ref{fig:Images123}). Other images probe smaller regions of the galaxy at very high magnification (e.g., images 1 and 2, Figure \ref{fig:Images123}). By studying the outflow properties in these different regions, one can ``zoom" into the same region of the galaxy at high spatial fidelity and quantify the variation of galactic outflow properties coming from the same region of the galaxy.

Figure \ref{fig:Images123}, bottom row, shows the pseudo-slits reconstructed in the source-plane. 
The far right square in the reconstructed source map of image 3 is probed twice in images 1 and 2 with different magnifications. This square contains regions U, E, and region B. For this square, we construct four pseudo-slits in image 1, and seven pseudo-slits in Image 2.

This effect is better visualized in Figure \ref{fig:3d_slits},  where the x-y plane represents the physical distance in the source-plane of the galaxy. The source-plane reconstructed image of the counter arc is shown for reference. The cyan horizontal planes represent the location of the three images of the galaxy: image 2 in the top plane, image 1 in the mid-plane, and image 3 in the bottom plane. The cyan color intensity of the planes of the 3 images in Figure \ref{fig:3d_slits} increases with increasing magnification of each image. The polygons show the location of each pseudo-slit in each image. Pseudo-slits with the same colors signify that the same physical region of the galaxy is being probed multiple times at different spatial sampling.

Figure \ref{fig:3d_slits} represents the top-down approach of this work, where we measure the average outflow property of this galaxy (from the counter arc,  area $\rm{\approx 60.5\ kpc^2}$ as measured from the star-forming region in the HST image), and zoom in to measure the local outflow properties at $\sim$ 20 -- 25 $\rm{kpc^2}$ (image 3) and  $\sim$ 0.5 -- 11 $\rm{kpc^2}$ (images 1 and 2) scales, respectively. We use the counter arc image as an indicator of the global values of the outflows of the galaxy. We use image 3 to give us the outflow properties at 3 distinct regions in the galaxy. Images 1 and 2 provide a detailed description of the local properties of the galaxy and the outflow.

\begin{figure*}
    \centering
    \includegraphics[width=\linewidth]{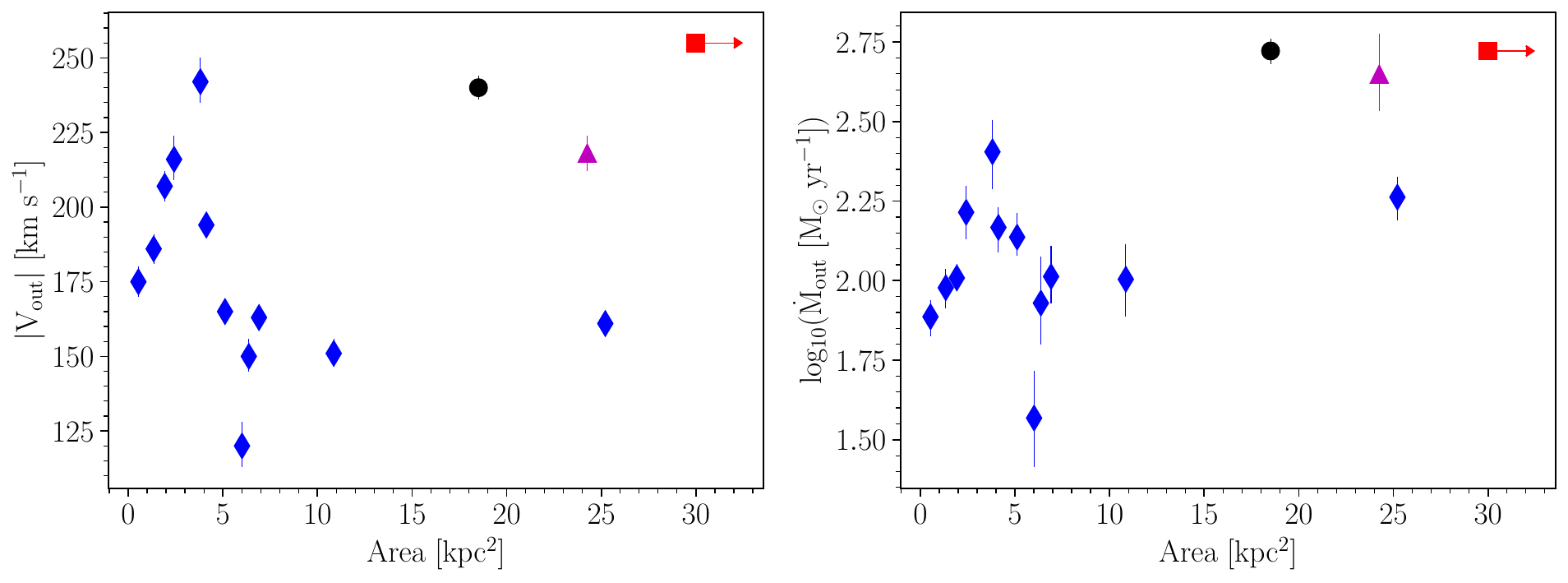}
    \caption{ Outflow properties as a function of area of pseudo-slits used to measure them. \textit{Left panel:} Outflow velocities as a function of area of pseudo-slits traced by the {\feii} transitions. Symbols of similar shape  and color represent the same physical region, that are  probed multiple times at different spatial scales in the source-plane. The colors of individual symbols are the same as Figure \ref{fig:3d_slits}. There is a variation in the outflow properties depending on the size of the area that we probe. The red square represents the values from the counter arc spectrum, with an area of 60.5 kpc$^2$, presenting a global measurement for the whole galaxy. \textit{Right panel:} Same as left panel, but now for mass outflow rates derived from {\feii} transitions. }
    \label{fig:Vout_Mout_Vs_Area_FeII}
\end{figure*}

Figure \ref{fig:Vout_Mout_Vs_Area_FeII} shows the effect of probing different areas in the source-plane on the measured outflow velocity (left panel) and mass outflow rates (right panel), respectively. We only show the measurements from the {\feii} transitions, as they are the most robust. The colors of different symbols correspond to the pseudo-slits of the same colors presented in Figure \ref{fig:3d_slits}. The symbols of different colors correspond to different physical regions of the galaxy, and the counter arc measurements are represented by red squares for the mean outflow velocity and mass outflow rates of the whole galaxy in Figure \ref{fig:Vout_Mout_Vs_Area_FeII}.

The blue diamonds in Figure \ref{fig:Vout_Mout_Vs_Area_FeII} correspond to the same physical region of that galaxy that are being studied at different spatial resolutions (from 25 $\rm{kpc^2}$ to 0.5 kpc$^2$ scales). This is similar to applying an adaptive mesh refinement in galaxy simulations, where one can zoom into the same part of the galaxy at higher spatial resolutions. The blue diamonds show a large spread in outflow velocities from $\rm{-120\ km\ s^{-1}}$ to $\rm{-242\ km\ s^{-1}}$, respectively. This spread becomes more prominent as we probe regions $<$ 6 kpc$^2$ in area. The mean outflow velocity for this whole region from these measurements is $\rm{-179\ km\ s^{-1}}$, which is slightly higher than the outflow velocity near the highest area probed at 25 $\rm{kpc^2}$ ($\rm{-161\ km\ s^{-1}}$). A similar spread is seen in the right panel for the mass outflow rates. The blue diamonds with area $\rm{< 12\ kpc^2}$ ranges from 37 {\MsolarPerYr} to 254 {\MsolarPerYr} with an average mass outflow rate $\approx$ 118 {\MsolarPerYr} and standard deviation $\approx$ 54 {\MsolarPerYr}. This mean value of the scatter is lower than the mass outflow rate of the blue diamond at 25 $\rm{kpc^2}$ (183 {\MsolarPerYr}). This shows the impact of the size and the local properties of the probed region on the outflow properties. Further, the traditional down-the-barrel approach of using integrated galaxy light to measure mass outflow rate may underestimate the local regions with extremely high mass outflow rates in a galaxy.

Furthermore, we calculate the mass outflow rates of the escaping outflowing gas, defined as the parts of the absorption lines where the velocity is greater than ${\rm V_{esc}}$. We use the apparent optical depth method \citep[AOD;][]{Savage1991AOD} to obtain the column density of the escaping fraction of the outflow gas. From this, we estimate the escaping mass outflow rate $\dot{M}_{esc}$ traced by {\mgii} 2796 and {\feii} 2382 for the different pseudos-slits. For the counter arc, we estimate $\dot{M}_{esc}$ to be $\mathrm{\approx 38_{-19}^{+18} M_{\odot}\ yr^{-1}}$ for {\mgii} 2796, which is $22_{-11}^{+11}\%$ of the total measured mass outflow rate. {\feii} 2382 gives an estimate of $\approx 17_{-9}^{+26}$ for $\dot{M}_{esc}$ in the counter arc, which is $3_{-2}^{+5}\%$ of the total $\dot{M}_{out}$ traced by {\feii}. This difference is due to the fact that {\mgii} 2796 has a higher oscillator strength ($f_{0}  = 0.6155$) compared to that of the {\feii} 2382 ($f_{0}  = 0.32$). The error bars are lower and upper limits on $\dot{M}_{esc}$. These limits are high due to the high error bars on the escape velocity ($V_{esc} \approx 418_{-89}^{+112} \mathrm{km\ s^{-1}}$) (See Table \ref{tab:mean_radius_outflow}). In addition, Region B almost does not show any escaping mass outflow rate. Typically, the outflowing gas that will escape the gravitational potential of this halo corresponds to $\sim$ 1 - 10\%  of the total mass outflow rates observed for the other pseudo-slits. We present all measurements of $\dot{M}_{esc}$ in Table \ref{table:M_P_E_dot}.

For the three independent regions with pseudo-slit areas of 20 -- 25 kpc$^2$ (Figure \ref{fig:Vout_Mout_Vs_Area_FeII}, blue, magenta and black points), the outflow velocities also vary between $\rm{-161\ km\ s^{-1}}$ and $\rm{-240\ km\ s^{-1}}$, respectively. These regions also show a variation in mass outflow rates of 183 {\MsolarPerYr} to 527 {\MsolarPerYr}. This shows large variations within the galaxy even at similar spatial resolution, suggesting that the outflow properties depend on the properties of the SF regions driving them.

These observations highlight the importance of understanding local properties of galaxies in driving galactic outflows. The average outflow properties, derived from the counter arc, fails to capture the factor of 2 variation in outflow velocities or the factor of 7 variation in mass outflow rates, in different regions of the same galaxy. When typical down-the-barrel experiments are conducted, where a single slit is placed in the brightest part of a galaxy, such observations might dramatically underestimate the local variations in the same galaxy. 

Given a robust lens model, this method can be employed for other gravitationally strong lensed galaxies at different redshifts. This will enable us in future to create a statistically large sample of spatially resolved measurements of galactic outflows and connect them directly back to their local driving sources.

\subsection{Comparison with Previous Work}

We compare the measured outflow velocities of {\rcsa} to the following outflow velocity measurements from the literature. The observations from {\rcsa} are multi-sightlines of one individual galaxy, while most of the literature is based on samples of different individual galaxies, and each galaxy has one integrated spectrum. We expect that the physics of the outflows in the early and late universe to be similar. The following studies cover a redshift range $z \sim$ 0.01--1.5, which is below the redshift of {\rcsa}, from: 
\begin{itemize} 
    \item \citet{bordoloi2014dependence}, where $\rm{V_{out}}$ is the mean velocity of the total {\mgii} absorption lines tracing the outflows in the stacked spectra of 486 star-forming galaxies in the redshift range $1\leq z \leq 1.5$.
    \item \citet{heckman2015systematic}, where $\mathrm{V_{out}}$ is the un-weighted mean velocity of the two UV absorption lines Si III 1206 {\AA}, N II 1084 {\AA} in a sample of local star-burst galaxies at $z<0.23$.
    
    \item \citet{chisholm2016shining}, where $\mathrm{V_{out}}$ is the equivalent width weighted velocity from Si II 1260 {\AA} in a sample of 37 star-forming and starburst galaxies at $z<0.26$.
    
    \item \citet{Xu2022CLASSYIII}, where $\mathrm{V_{out}}$ is the central velocity of the ouflowing Gaussian component from Si II and Si IV in a sample of local star-burst galaxies at $z<0.18$.
\end{itemize}

\begin{figure*}
    \centering
    \includegraphics[width=0.8\textwidth]{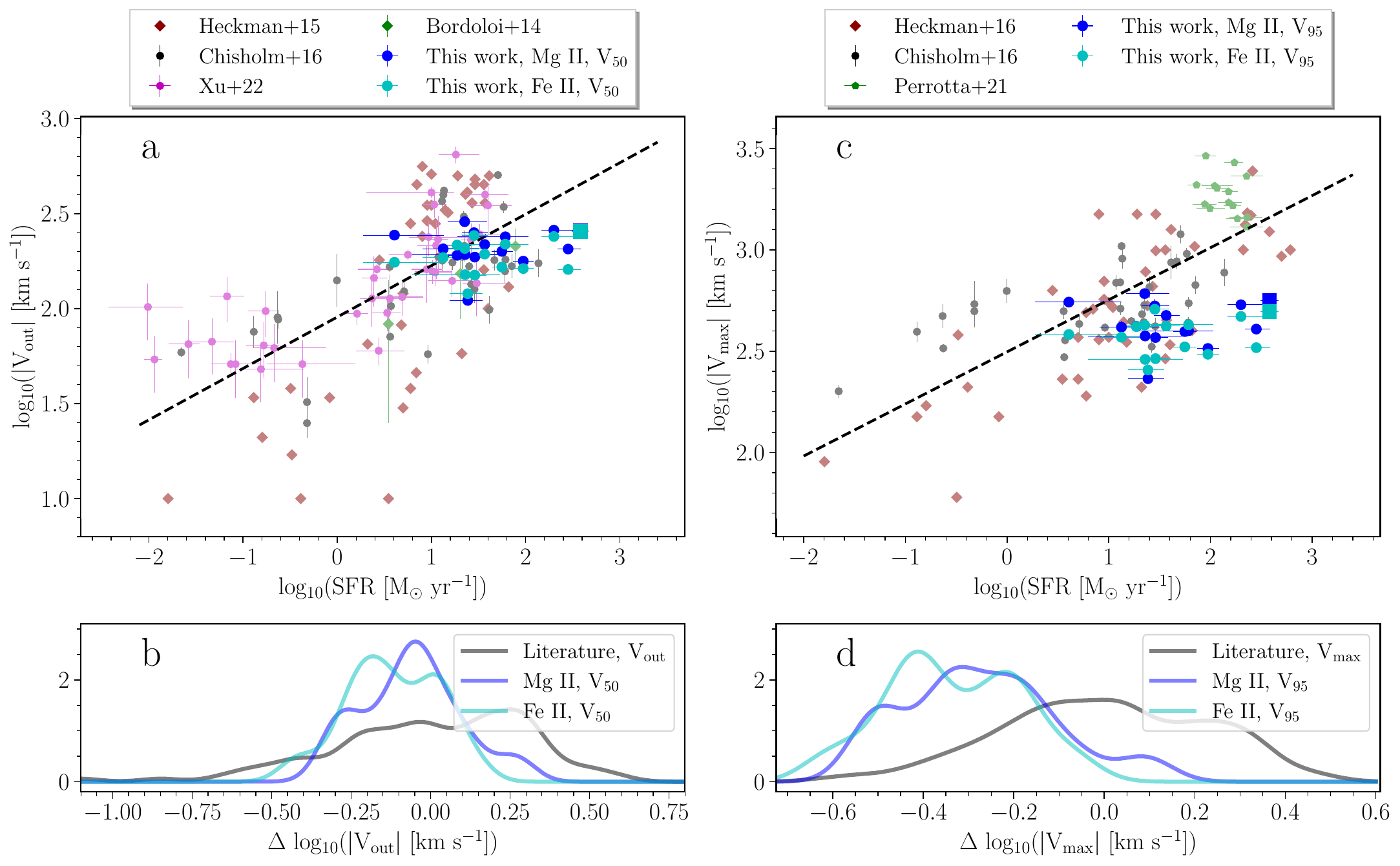}
    \caption{\textit{Panel (a):} Variation of the $\mathrm{50^{th}}$ percentile outflow velocity with SFR. The blue and cyan circles represent the outflow velocities traced by the {\mgii} and {\feii} for {\rcsa}, respectively. The cyan and blue square points are for the measurements of the counter arc for {\feii} and {\mgii}, respectively.  We compare this work with the literature measurements from \citet{bordoloi2014dependence} (green diamonds), \citet{heckman2015systematic} (brown diamonds), \citet{chisholm2016shining} (gray circles), and \citet{Xu2022CLASSYIII} (magenta circles). The black dashed line represents the best-fit line for the literature measurements. \textit{Panel (b):} Kernel density estimates (KDE) for the residual outflow velocities ($\mathrm{\Delta \log_{10}(|V_{out}|) = \log_{10}(|V_{50}|)  - \log_{10}(|V_{50, literature, fit}|)}$) for the {\mgii} (blue curve), {\feii} (cyan curve) from {\rcsa}, and the literature (black curve) measurements.  \textit{Panel (c):} Maximum outflow velocity dependence on SFR. The blue and cyan points represent the $\mathrm{95^{th}}$ percentile outflow velocities traced by the {\mgii} and {\feii} for {\rcsa}, respectively. The square points are for the measurements of the counter arc. We compare these measurements with the literature measurements from \citet{heckman2016implications} (brown diamonds; bluest part of the absorption tail velocity), \citet{chisholm2016shining} (gray circles; $\mathrm{90^{th}}$ percentile Si II velocity), and \citet{Perrotta2021ExtStarburstOutflows} (green hexagons; $\mathrm{98^{th}}$ percentile {\mgii} velocity). The dashed straight black line is the best-fit line for the literature data points. \textit{Panel (d):} KDE for the residual maximum outflow velocity ($\mathrm{\Delta \log_{10}(|V_{max}|) = \log_{10}(|V_{max}|)  - \log_{10}(|V_{max, literature, fit})|}$) with the same color scheme as panel (b). }
    \label{fig:SFR_Vout_Vmax}
\end{figure*}

Figure \ref{fig:SFR_Vout_Vmax}, panel (a) shows the {\feii} (cyan) and {\mgii} (blue) outflow velocity measurement from {\rcsa}, and the measurements from the literature as a function of SFR. The SFR measurements in {\rcsa} are also spatially resolved, as mentioned in {\S} \ref{sec:GalaxyProperties}. The outflow velocities measured in individual regions of {\rcsa} follow the broad trend of increasing outflow velocity with increasing SFR. However, a large scatter is seen in outflow velocities of {\rcsa} which is comparable to that seen in local star-burst galaxies. To quantify this scatter, we fit a straight line to all the literature measurements (dashed black line). The integrated outflow velocity value from the counter arc is offset from the best-fit line from the literature points by $\approx$ -0.25  dex for both {\feii} and {\mgii} measurements. We subtract this mean profile from each individual measurement and compute the residual velocity differences, which takes out any trend of outflow velocities with SFRs. Figure \ref{fig:SFR_Vout_Vmax}, panel (b) shows the  the kernel density estimate (KDE) of these residual outflow velocity distributions  for the literature data (black curve), {\feii} measurements for this work (cyan curve), and {\mgii} measurements for this work (blue curve), respectively. For {\rcsa}, the {\feii} and {\mgii} measurements are below the best fit line from the literature measurements, and medians of their residual distributions are offset from the literature values by -0.1 dex and -0.2 dex for {\mgii} and {\feii}, respectively. From {\rcsa}, the scatter has a width of $\approx$ 0.9 dex, which is slightly smaller than the width of the scatter of the literature data points that has a width of $\approx 1.5$ dex. 

We perform the same analysis for the maximum outflow velocity $\mathrm{V_{max}}$. For {\rcsa}, this corresponds to the $\mathrm{95^{th}}$ percentile velocity ($V_{95}$) of the outflow component of {\feii} and {\mgii} absorption lines. We compare these measurements with that from: 
\begin{itemize}
    \item \citet{heckman2016implications}, where $\mathrm{V_{max}}$ is the velocity of the bluest part of the absorption lines Si II 1260{\AA}, C II 1334{\AA}, and C II 1036{\AA} before it reaches the continuum from a sample of extreme star-burst galaxies at $0.4\leq z \leq 0.7$.
    \item \citet{chisholm2016shining}, where $\mathrm{V_{max}}$ is the outflow velocity of the Si II 1260 {\AA} line at the 90 percent of the continuum of the same sample of galaxies in the previous paragraph.
    \item \citet{Perrotta2021ExtStarburstOutflows}, where $\mathrm{V_{max}}$ is the 98$\rm{^{th}}$ percentile velocity for {\mgii} in a sample of massive compact starburst galaxies at $z\sim 0.4 - 0.7$.
\end{itemize}

We caution the reader that different literature measurements use different definition of $\mathrm{V_{max}}$, as described above. To simplify comparing these new observations with literature measurements,we adopt $V_{95}$ as the maximum outflow velocity. We plot $\mathrm{V_{max}}$ as a function of SFR in Figure \ref{fig:SFR_Vout_Vmax}, panel (c). The straight dashed line is the best-fit line of the literature measurements. The majority of the measurements for the {\feii} and {\mgii} from {\rcsa} are similar to the scatter from the literature sample. However, the {\rcsa} points are below the best-fit line. The {\feii} and {\mgii} maximum outflow velocities from the counter arc are offset from the best-fit line of the literature data points by $\approx$ -0.46 dex and -0.41 dex, respectively. We subtract the straight line from all the measurements to obtain the residual maximum outflow velocities. Figure \ref{fig:SFR_Vout_Vmax}, panel (d), shows the KDEs for the residual outflow velocities of the literature measurements (black curve), {\feii} (cyan curve), and {\mgii} (blue curve), respectively. These distributions show a similar trend to what is observed for the median outflow velocity. The {\feii} and {\mgii} KDEs medians are skewed to the negative side of the peak of the literature distribution by -0.3 dex and -0.2 dex for {\feii} and {\mgii}, respectively. In one galaxy, the width of the scatter or total scatter is $\approx$ 0.9 dex, which is similar to the total scatter of the literature data points from many galaxies at the lower redshifts that is $\approx 1.05$ dex.

\begin{figure*}
    \centering
    \includegraphics[width=0.8\textwidth]{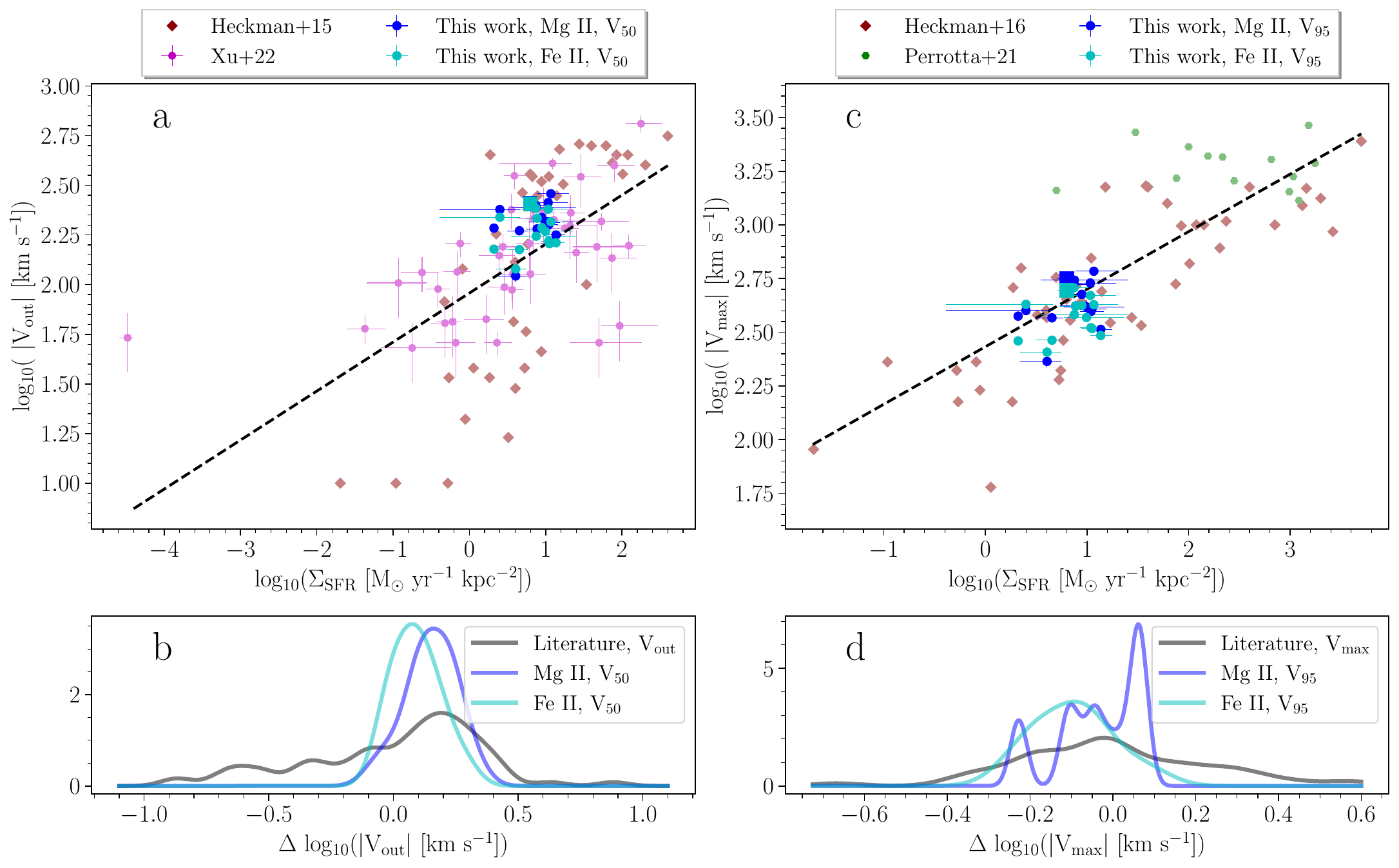}
    \caption{ \textit{Panel (a):} $\mathrm{50^{th}}$ percentile outflow velocity versus SFR per unit area ($\Sigma_{SFR}$). The blue and cyan circles represent the outflow velocity for {\mgii} and {\feii} in {\rcsa}, respectively. The cyan and blue squares are the average measurements of the counter arc for {\feii} and {\mgii}, respectively. The literature data points in this panel are from \citet{heckman2015systematic} (brown diamonds), and \citet{Xu2022CLASSYIII} (magenta diamonds). The black dashed line is the best fit line for the literature data points. \textit{Panel (b):} KDE  of the residual outflow velocity with respect to the literature best fit line $(\rm{\Delta \log_{10}(|V_{50}|) = \log_{10}(|V_{50}|) - \log_{10}(|V_{50,\ best-fit}|)})$ for {\mgii} (blue curve), {\feii} (cyan curve) from {\rcsa}, and the literature measurements (black curve). \textit{Panel (c):} Maximum outflow velocity versus $\Sigma_{SFR}$. The blue and cyan circles are for the {\mgii} and {\feii} for {\rcsa}, respectively. The literature measurements are from \citet{heckman2016implications} (brown diamonds), and \citet{Perrotta2021ExtStarburstOutflows} (green hexagons). The maximum velocities here have the same definition as in Figure \ref{fig:SFR_Vout_Vmax}. \textit{Panel (d):} KDE of the residual maximum outflow velocity with respect to the literature best fit line $(\rm{\Delta \log_{10}(|V_{max}|) = \log_{10}(|V_{max}|) - \log_{10}(|V_{max,\ best-fit})|})$ with the same color scheme panel (b). }
    \label{fig:SigmaSFR_V50_V95}
\end{figure*}

From Figure \ref{fig:SFR_Vout_Vmax}, it is clear that in {\rcsa}, the regions with highest SFRs exhibit slower outflow velocities, compared to the local sample of star-burst galaxies. This might be owing to the fact that higher SFR is observed in pseudo-slits with larger areas. We correct for this effect by performing the comparison of outflow velocity with star-formation rate surface density ($\mathrm{\Sigma_{SFR}}$) of individual pseudo-slits. Panel (a) of Figure \ref{fig:SigmaSFR_V50_V95} shows how the median outflow velocity depend on $\mathrm{\Sigma_{SFR}}$. The {\mgii} (blue circles) and {\feii} (cyan circles) measurements and the literature measurements from \citet{heckman2015systematic} (brown diamonds), and \citet{Xu2022CLASSYIII} (magenta circles) are shown in this panel. The dashed black line represents the best-fit line between $\log(|V_{out}|)$ and $\log (\Sigma_{SFR})$ for the literature data points. The offset of the integrated median outflow velocity from the counter arc is offset from the best-fit line of the literature data points by $\approx$ 0.25 dex for both {\feii} and {\mgii}. In order to quantify the scatter, we subtract our measurements and the literature measurements from the best fit line and produce KDEs from the residual outflow velocities in panel (b) for {\feii} (cyan curve), and {\mgii} (blue curve), and literature (black curve). The scatter in the literature points covers $\approx$ 1.5 dex. While both the scatters in {\feii} and {\mgii} have range $\approx$ 0.6 dex, which is less than half the scatter of the literature. The median values of the residual velocity distributions are almost identical within $0.05$ dex. This shows that in the $\Sigma_{SSFR}$ space, the outflow velocities of {\rcsa} are similar to what is seen for the local population of star-burst galaxies. 
Panel (c) of Figure \ref{fig:SigmaSFR_V50_V95} shows the relation between $V_{max}$ and $\Sigma_{SFR}$, the cyan and blue represent the {\feii} and {\mgii} measurements for RCS0327, respectively. The brown diamonds and green hexagons are from \citet{heckman2016implications} and \citet{Perrotta2021ExtStarburstOutflows}, respectively. The dashed black line is the best fit line for the literature data points. The integrated maximum outflow velocity for both {\feii} and {\mgii} from the counter arc are offset from the best-fit line of the literature by 0.05 and 0.1 dex, respectively. Similar to the left column of the Figure, we produce the KDEs of residual velocity distributions of {\rcsa} and the literature measurements in panel (d) of Figure \ref{fig:SigmaSFR_V50_V95}. The ranges of the scatter for {\feii} and {\mgii} are $\sim$ 0.55 dex and 0.45 dex, respectively, which are almost less than half the scatter of the literature ($\sim 1 $dex). The medians of the three distributions  are $\sim 0$. This shows that there is no apparent trend for {\rcsa} velocity measurements with $\Sigma_{SFR}$ for either the individual star-forming regions at different physical areas or the total integrated measurements from the counter arc.

These results show that the outflow kinematics from one single lensed galaxy are comparable to the outflow kinematics of a large sample of star-burst galaxies within the redshift range $0 \leq z \leq 1.5$. This highlights the large variations in outflow kinematics observed within a single galaxy.

\subsection{Outflow Momentum Flux and Energy Flux}
From the mass outflow rates and the outflow velocity, we can calculate the momentum flux ($\dot{p}_{out}$) and energy flux ($\dot{E}_{out}$) carried by the outflows. We will use the median outflow velocity ($V_{out}$), where most of the absorption is located along the line of sight. This can be calculated using the following equations:
\begin{equation}
    \dot{p}_{out} = \dot{M}_{out}\ V_{out},
\end{equation}

\begin{equation}
    \dot{E}_{out} = \frac{1}{2} \dot{M}_{out}\ V_{out}^2.
\end{equation}

These measurements are summarized in Table \ref{table:M_P_E_dot} in Appendix \ref{appendix:Outflow_Properties}. We compare them with the momentum and energy fluxes supplied by the starburst and assess whether the outflows are energy-driven or momentum-driven across the different sizes of the regions that we probe in the different images.

\begin{figure*}
    \centering
    \includegraphics[width=0.8\textwidth]{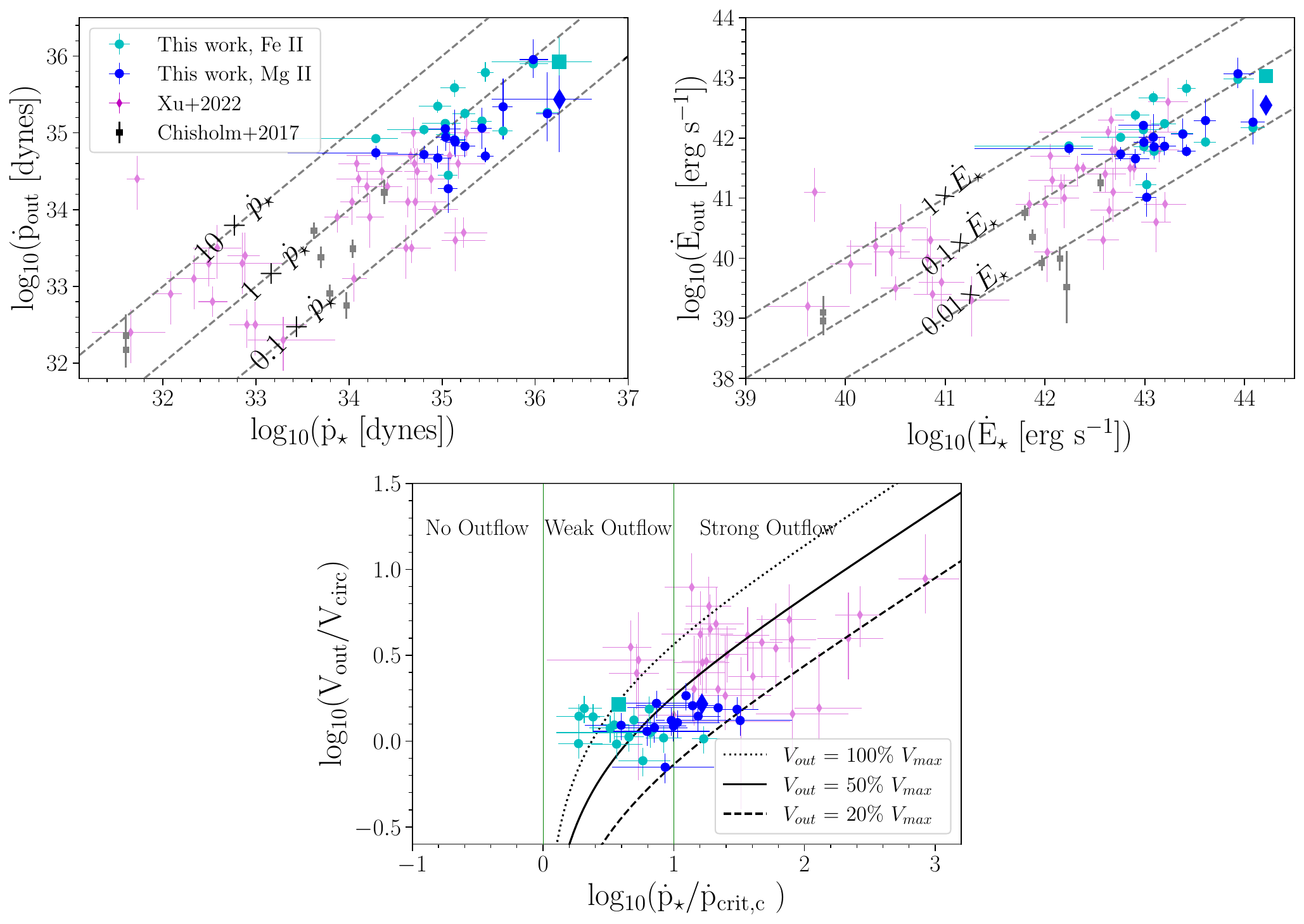}
    \caption{\textit{Top left:} Outflow momentum flux versus momentum flux from the star-formation activities. 
    \textit{Top right:} Outflow energy flux versus the stellar energy flux. The cyan and blue circles represent the measurements of the individual regions of the main arc from {\feii} and {\mgii}, respectively. The cyan square and blue diamond represent the {\feii} and {\mgii} measurements from the counter arc, respectively. The magenta diamonds and gray squares are the measurements from \citet{Xu2022CLASSYIII} and \citet{chisholm2017GalacticOutflows}, respectively. The three diagonal solid gray lines represent the ratio of $\dot{E}_{out}/\dot{E}_{\star}$ with values 0.01, 0.1 and 1. The majority of the data points are below $0.1 \times \dot{E}_{\star}$. Our data points for energy and momentum flux are consistent with measurements from \citet{Xu2022CLASSYIII} and \citet{chisholm2017GalacticOutflows}. \textit{Bottom:} Normalized outflow velocity ($\mathrm{v_{out}/v_{cir}}$) versus the ratio of the stellar momentum flux to the critical momentum flux value ($\dot{p}_{\star}/ \dot{p}_{crit, c}$) from \citet{heckman2015systematic}. We also plot the three outflow regimes (No outflow, Weak outflow, and Strong outflow). Most of the {\feii} data points are within the weak outflow regimes. While the {\mgii} data points are divided between these two regimes. The \textit{dotted, solid}, and \textit{dashed} curved lines represent the predicted outflow velocity as a fraction of the $V_{max}$, $\mathrm{V_{out}}$ = 100\%, 50\%, and 20\% $\times V_{max}$, respectively. }
    \label{fig:P_flux_V_norm_Rc_Eflux}
\end{figure*}

From \texttt{Starburst99} models \citep{leitherer1999starburst99} associated with Supernovae, we can calculate the stellar momentum flux ($\dot{p}_{\star}\ [dynes] = 4.6 \times 10^{33}\  \mathrm{SFR\  [M_{\odot}\ yr^{-1}]}$) and energy flux ($\dot{E}_{\star}\ [erg\ s^{-1}] = 4.3 \times 10^{41}\  \mathrm{SFR\ [M_{\odot}\ yr^{-1}] }$) provided to the outflows from the star formation activities in our selected regions \citep{Xu2022CLASSYIII}. Figure \ref{fig:P_flux_V_norm_Rc_Eflux}, top left panel, shows the $\dot{p}_{out}$ from the measured outflow component of the absorption lines versus the stellar momentum flux. The three diagonal solid lines in the panel represent the ratio of $\dot{p}_{out} / \dot{p}_{\star}$ with values 0.1, 1, and 10. We see that the outflow momentum flux is at least 10\% of the stellar momentum flux for all measurements in {\rcsa}. Furthermore, the majority of the data points are around 100\% of $\dot{p}_{\star}$, with some of the data points above 100\% of $\dot{p}_{\star}$. For the outflow observed in this galaxy to be momentum driven, the transfer of momentum from star-formation to driving outflow must be 100\% efficient. It is also possible that the current models are not adequately reflecting the momentum injected by the stellar populations.

The top right panel of Figure \ref{fig:P_flux_V_norm_Rc_Eflux} shows the $\dot{E}_{out}$ traced by {\feii} and {\mgii} versus the energy flux $\dot{E}_{\star}$ provided by the star-formation activities in the galaxy. The three diagonal lines in this panel represent the ratio of $\dot{E}_{out} / \dot{E}_{\star}$ with values 0.01, 0.1, and 1, respectively. Most of the measurements in {\rcsa } from {\mgii} and about half of the measurements from {\feii} are below 0.1.
These estimates suggest that most of the outflowing gas in {\rcsa} can be driven by  $\sim$ 10\% of the energy generated in star-formation. Comparing the two scenarios of momentum and energy flux estimates suggest that energetically, it is more feasible to transfer 10\% of the energy flux into driving a galactic outflow than transferring 100\% of the momentum flux. This consideration leads us to suggest that the cool outflow observed in {\rcsa} is an energy-driven outflow.

In addition, the scatter in both panels in momentum and energy fluxes values that are measured from one star-forming galaxy {\rcsa} at $ z \approx 1.7$ is similar to the scatter from a large sample of star forming galaxies in the local universe $z < 0.2$ from \citet{Xu2022CLASSYIII} shown as magenta diamonds in both of the top panels in Figure \ref{fig:P_flux_V_norm_Rc_Eflux}, and is also similar to the scatter of the measurements from \citet{chisholm2017GalacticOutflows}, shown as gray squares in Figure \ref{fig:P_flux_V_norm_Rc_Eflux}. This further highlights that in a single galaxy (\rcsa), the variation in energy and momentum fluxes are as diverse as observed in a population of local star-burst galaxies.

\subsection{Comparison of Observations and Theory}\label{sec:compare_obs_theory}

\citet{heckman2015systematic} introduced a model for outflowing clouds under the effect of the inward force of gravity, and an outward force due to the momentum flux supplied to the outflow by the starburst. The inward force due to gravity on a cloud located at radius $r$ from the star forming region can be expressed as 
\begin{equation}\label{eq:heckmanInForce}
    F_{in} = M_c \frac{v_{cir}^2}{r},
\end{equation}
where $M_c$ is the mass of the cloud, and $v_{cir}$ is the circular velocity ($v_{cir} = \sqrt{G\frac{M(<r)}{r} }$, $M(<r)$ is the mass of the galaxy enclosed within radius $r$). The mass of the cloud can be written as: $M_c = A_c\ N_c\mu\ m_p$, where $A_c$ is the cross-sectional area of the cloud, $N_c$ is the hydrogen column density of the cloud, and $\mu\ m_p$ is the mean mass of neutral hydrogen. The outward force acting on the cloud can be written as: 
\begin{equation}\label{eq:heckmanOutForce}
    F_{out} = \frac{A_c}{4\pi r^2} \dot{p}_{\star},
\end{equation}
where $\dot{p}_{\star}$ is the momentum flux provided to the cloud by the star-formation activities. At equilibrium, we can calculate the critical momentum flux at which a cloud at the boundary of the star forming region ($r = r_{\star}$) will exactly balance gravity. This critical momentum flux is: 
\begin{equation}\label{eq:CritPdotHeckman}
    \dot{p}_{crit} = 4\pi\ r_{\star} N_c\mu\ m_p\ v_{cir}^2
\end{equation}

Figure \ref{fig:P_flux_V_norm_Rc_Eflux}, bottom panel, shows the normalized outflow velocity ($v_{out}/v_{cir}$) versus the ratio of the stellar momentum flux and the critical momentum flux ($R_{crit} = \dot{p}_{\star}/ \dot{p}_{crit}$) as three black lines from this model, where the predicted outflow velocity ($V_{out, model}$) is 20\%, 50\%, and 100\% of the maximum velocity of the model  ($V_{max, model}$). The blue and cyan points are {\mgii} and {\feii} traced outflow measurements in {\rcsa}, and the magenta diamonds are measurements of local star-burst galaxies from \citet{Xu2022CLASSYIII}, respectively. Most of the measurements for {\rcsa}  are between 20\% and 100\% of the maximum velocity of the model.

We also show three regimes for $R_{crit}$: (1) $R_{crit} < 1$ (the stellar momentum flux provided to the outflow is less than the critical momentum flux; this leads to no outflow), (2) $1 < R_{crit} < 10$ (where the stellar momentum flux is at least equal to the critical momentum or up to 10 times greater. This is the weak outflow regime), (3) $R_{crit} > 10$ (where the stellar momentum flux is at least 10 times larger than the critical momentum flux. This is the strong outflow regime). Figure \ref{fig:P_flux_V_norm_Rc_Eflux}, bottom panel, shows that all our selected regions (blue and cyan points) in RCS0327 are hosting outflows. Most of the {\feii} data points are within the weak outflows regimes, except one data point in the strong outflow regime. However, the majority of {\mgii} measurements are in the strong outflow regime. Note that {\mgii} absorption is  saturated in this work, and hence the inferred total hydrogen column densities are highly uncertain and underestimated. We present the {\mgii} $R_{crit}$ values for completeness but focus on the inferred $R_{crit}$ from the {\feii} transition for interpretation.

The differences between measurements of individual regions and the global values from the counter arc highlight the importance of the stellar properties of the local star-forming individual regions, which drive these outflows.

\section{Summary \& Conclusion}\label{sec:conclusions}
In this work, we study the cool outflows in the gravitationally lensed galaxy {\rcsa} at $z= 1.70347$ with the VLT/MUSE IFU observations using multiple pseudo-slits in the image-plane in 4 multiple images due to lensing. These 15 pseudo-slits correspond to 15 distinct spatial locations on the sky, each probing different magnifications (2--100) and physical areas (0.5 kpc$^2$ -- 25 kpc$^2$) of the same galaxy. This produces the outflow properties over different spatial resolutions, and we study how they are related to the star forming regions that may be driving them. These measurements will be helpful for numerical simulations that consider the effect of mesh refinement on the implementation of outflows as a feedback mechanism. 
We summarize as follows:
\begin{itemize}
    \item {\rcsa} exhibits average outflow  velocities of $\approx -254$ to $-257\ \rm{km\ s^{-1}}$ traced by both the {\feii} and {\mgii} absorption lines, respectively. These measurements are from the blueshifted absorption lines seen in the counter arc spectrum, which represents the average outflow velocity for the whole galaxy (Figures \ref{fig:FeII_fitting}, \ref{fig:MgII_fitting}, and \ref{fig:corner_measured_outflows}). 
    
    \item We detect {\mgii} resonant and {\feiiems} fine-structure emission lines in all individual local regions of \rcsa. We do not detect any {\feii} resonant emission lines. The {\mgii} resonant emission lines are redshifted ($\rm{ V_{ems} \approx}$ 15 -- 165 $\rm{\ km\ s^{-1}}$), consistent with the interpretation that this emission is arising via scattering of outflowing {\mgii} photons from the back-side of the outflowing gas. The {\feiiems} emission lines are consistent with being around the systemic redshift of the galaxy in the velocity range $\sim -73$  to 25 $\mathrm{km\ s^{-1}}$ (Figures \ref{fig:FeII_fitting}, \ref{fig:MgII_fitting}, and \ref{fig:emission_vel_MgII_FeII}).

    \item We detect a secondary {\mgii} resonant emission lines in 12 out of the 15 individual spectra of \rcsa, with $\rm{ V_{ems} \approx}$ 372--679 $\rm{km\ s^{-1}}$. This line is kinematically offset from the primary {\mgii} emission lines by $\rm{\Delta V_{ems} \approx 400\ km\ s^{-1}}$. We interpret the presence of this second line as probing a secondary outflowing shell of gas, perhaps tracing an earlier episode of star-burst in {\rcsa} (Figures \ref{fig:MgII_fitting} and \ref{fig:emission_vel_MgII_FeII}).

    \item We measure the mass loading factor ($\eta = \dot{M}_{out} / SFR$) at individual local regions of {\rcsa}. Within the same galaxy, $\eta$ varies by an order of magnitude ($\eta \sim$ 1 -- 10). The counter arc, which represents the average light from the entire galaxy has $\eta \approx 1$, significantly less than that measured in the majority of individual local regions of {\rcsa}. This suggests that the global measurement from the counter arc smears the measurements from local star-forming regions (Figure \ref{fig:Mdot_out_SFR_SSFR}).

    \item We investigate the small scale variation of outflow velocities and mass outflow rates of the same region of {\rcsa} galaxy, probed at different spatial scales. This is possible, as {\rcsa} is multiply imaged by graviational lensing, and different images have different magnifications. We probe the same region (Image 3.1, See Appendix \ref{appendix:Outflow_Properties}) of the galaxy with the pseudo-slit sizes of 0.5 kpc$^2$ to 25 kpc$^2$, where outflow velocities vary between ($\rm{-120\ km\ s^{-1}}$ to $\rm{-242\ km\ s^{-1}}$). The average outflow velocity of the region (probed by the largest pseudo-slit of area 25 kpc$^2$) is $\rm{-161\ km\ s^{-1}}$. This shows that small scale variation in outflow kinematics is smeared out in any average down-the-barrel observations (Figures \ref{fig:MgII_FeII_source} and \ref{fig:Vout_Mout_Vs_Area_FeII}). 

    \item For the same regions, mass outflow rates vary between 37 and 254 {\MsolarPerYr}. The largest pseudo-slit shows a mass outflow rate of 183 {\MsolarPerYr}. These measurements highlight the importance of spatially resolved observations in interpreting the galactic outflow properties from down-the-barrel observations (Figure \ref{fig:Vout_Mout_Vs_Area_FeII}). 

    \item For three physically independent regions that are probed with pseudo-slits of comparable size (20 -- 25 kpc$^2$), outflow velocities vary between $-161\ \rm{km\ s^{-1}}$ and $-240\ \rm{km\ s^{-1}}$ (Figure \ref{fig:Vout_Mout_Vs_Area_FeII}).
    
    \item These regions exhibit a variation in mass outflow rates between $183_{-28}^{+29}$ and $527_{-47}^{+48}$ {\MsolarPerYr}. The global mass outflow rate from the counter arc is $527_{-29}^{+30}$ {\MsolarPerYr} (Figure \ref{fig:Vout_Mout_Vs_Area_FeII}). 

    \item The scatter in the outflow velocity measurements from one lensed galaxy at $z\sim 1.7$ ($\rm{V_{out} \sim -120\ to -255\ km\ s^{-1} }$) is comparable to large samples of galaxies at lower redshifts (Figures \ref{fig:SFR_Vout_Vmax} and \ref{fig:SigmaSFR_V50_V95}).

    \item We compare the momentum and energy fluxes associated with the outflowing gas from {\rcsa} to the momentum and energy fluxes provided by the star-formation activity in this galaxy. The outflow momentum flux in this galaxy is $\gtrsim$ 100\% of the momentum flux provided by star-formation. Whereas, the outflow energy flux is $\approx$ 10\% of the total energy flux provided by star-formation. These estimates suggest that the outflow in {\rcsa} is energy-driven or that the stellar models significantly underestimate the momentum flux provided by stellar populations (Figure \ref{fig:P_flux_V_norm_Rc_Eflux}).   

    \item The outflow measurements from this galaxy are consistent with the analytical model of outflow cloud under the effect of the star-formation activity and gravity from \citet{heckman2015systematic} (Figure \ref{fig:P_flux_V_norm_Rc_Eflux}). 
    
\end{itemize}

The method that we use in this paper for studying the outflows can be applied to more gravitationlly lensed galaxies with multiple images in the image-plane. This will provide a large sample of spatially resolved outflows measurements to guide and constraint current theoretical models and numerical simulations of galaxy evolution. Other ground-based IFUs like the Cosmic Web Imager (KCWI) at the Keck telescopes \citep{morrissey2018keck} or space-based IFUs like Near Infrared Spectrograph (NIRSpec) \citep{Jakobsen2022NIRSpec, Boeker2023NIRSpec} on \textit{James Webb Space Telescope} (JWST) will provide us with observations of such galaxies.

\section*{Acknowledgements}
This material is based upon work supported by the National Science Foundation under Grant No. NSF-AST 2206853. This work is based on observations collected at the European Organization for Astronomical Research in the Southern Hemisphere under ESO program 098.A-0459(A). In addition, we used observations made with the NASA/ESA Hubble Space Telescope, obtained from the data archive at the Space Telescope Science Institute (STScI). STScI is operated by the Association of Universities for Research in Astronomy, Inc. under NASA contract NAS 5-26555. L.F.B. acknowledges support from ANID BASAL project FB210003 and FONDECYT grant 1230231. S.L. acknowledges support by FONDECYT grant 1231187.

We use the python software packages \texttt{astropy} \citep{astropy2013astropy,astropy2018astropy}, \texttt{matplotlib} \citep{Hunter2007Matplotlib}, and \texttt{scikit-learn} \citep{scikitlearn2011}.

\section*{Data Availability}
The MUSE data is available for the Public through the European Southern Observatory web page. The rest of the data products are available from the corresponding authors based on a reasonable request.

\bibliographystyle{mnras}
\bibliography{myBibliography} 



\appendix
\section{ }
\subsection{Escape Velocity}\label{appendix:escape_velocity}
We compute the escape velocities $V_{esc}$ of the galaxy at different R by assuming that the density of the dark matter halo of the galaxy follows an NFW profile \citep{navarro1996NFW}. We define the escape velocity at radius R as $V_{esc} ( V_{esc}(R) = \sqrt{2|\phi(R)|} )$, where $\phi(R)$ the gravitational potential well of the dark matter halo of this galaxy \citep{bryan1998XrayClusters, Klypin2002DynamicalModels, dutton2014CDMHaloes}.

In more detail, the NFW potential $\phi(R)$ can be written as:
\begin{equation}
    \phi(R) = \frac{GM_{vir}}{f(c_{vir})} ln\left(1 + \frac{R}{\kappa}\right) \frac{1}{R}, 
\end{equation}
where $M_{vir}$ is the virial mass, $c_{vir}$ is the concentration parameter that is the ratio between the virial radius $R_{vir}$ and the scale radius of the NFW profile, and $\kappa$ is the ratio between the virial radius $R_{vir}$ and the concentration parameter $c_{vir}$. The function $f(c_{vir})$ is
\begin{equation}
    f(c_{vir}) = ln(1+c_{vir}) - \frac{c_{vir}}{1 + c_{vir}}
\end{equation}
The concentration parameter can be calculated using the redshift of the galaxy and the virial mass of the halo as described in \citet{dutton2014CDMHaloes}:
\begin{equation}
    \mathrm{log}(c_{vir}) = a + b\ \mathrm{log}\left(\frac{M_{vir}}{10^{12} M_{\odot}}\right), 
\end{equation}
where the parameters $a$ and $b$ are defined as:
\begin{equation}
    a = -0.097 + 0.024 z    
\end{equation}

\begin{equation}
    b = 0.537 + (1.025 - 0.537)\ exp(-0.718 z^{1.08})
\end{equation}

The virial radius $R_{vir}$ can be expressed in terms of $M_{vir}$ using the relation of the mean halo mass density:
\begin{equation}
    \langle\rho_{halo}\rangle = \frac{3 M_{vir}}{4\pi R_{vir}^3} = \Delta_{c} (z) \Omega_m(z) \frac{3H^2(z)}{8 \pi G},
\end{equation}
where $\Omega_m(z)$ is the matter density in the universe as a function of $z$, and the parameter $\Delta_c(z)$ is function of $\Omega_m(z)$. From \citet{bryan1998XrayClusters}, the parameter $\Delta_{c}(z)$ can be expressed as:
\begin{equation}
    \Delta_c(z) = 18 \pi^2 + 82 x - 39 x^2,
\end{equation}
where $x = \Omega_m(z) - 1$. 
The calculated values for $\mathrm{V_{esc}}$ at different $R$s are summarized in Table \ref{tab:mean_radius_outflow}.

\subsection{Reported Outflow Measurements}\label{appendix:Outflow_Properties}
In this appendix, we summarize the measured and calculated outflow properties for both the absorption lines and emission lines for the selected pseudo-slits. 
Table \ref{tab:mean_radius_outflow} represents the measured weighted outflow radii $\langle R \rangle$ from the {\feii} and {\mgii} surface brightness radial profiles from \citetalias{Shaban202230kpc}.

Table \ref{table:outflow_properties} shows the rest-frame equivalent width $\mathrm{W_r}$, outflow velocity $\mathrm{V_{out}}$, covering fraction $C_f$, column densities $\mathrm{N}$, the 90$\rm{^{th}}$ percentile outflow velocity $V_{90}$, and 95$\rm{^{th}}$ percentile outflow velocity $V_{95}$ for the outflow component for the {\feii} and {\mgii} absorption lines. 

Table \ref{table:M_P_E_dot} represents the measured mass outflow rates $\dot{M}_{out}$, outflow momentum flux $\dot{p}_{out}$, and outflow energy flux $\dot{E}_{out}$ for the {\mgii} and {\feii} absorption lines. 

Table \ref{tab:outflows_emission} represents the emission lines' best-fit model parameters and measured properties: rest-frame equivalent width $\mathrm{W_r}$, emission velocity $\mathrm{V_{ems}}$, and Doppler parameter $\mathrm{b_D}$ for the {\mgii} primary and secondary emission components, and the {\feiiems} emission.

\begin{table}
    \centering
    \begin{tabular}{ccccc}
        \hline\hline 
        Region & $\langle R \rangle _{\feii}$ & $\langle R \rangle _{\mgii}$ & $\rm{V_{esc, \feii}}$ & $\rm{V_{esc, \mgii}}$\\
        Name & [kpc] & [kpc] & $\mathrm{km\ s^{-1}}$ & $\mathrm{km\ s^{-1}}$\\
        (1) & (2) & (3) & (4) & (5)\\
        \hline
        knot E & 4.7 & 6.05 & $419_{-89}^{+112}$ & $414_{-89}^{+112}$ \\ 
        knot U & 4.76 & 5.29 & $419_{-89}^{+112}$ & $417_{-89}^{+112}$ \\ 
        knot B & 5.66 & 6.7 & $416_{- 89 }^{+112}$ & $412_{-89}^{+112}$ \\ 
        Image 3 & 9.66 & 10.64 & $403_{-89}^{+112}$ & $400_{-89}^{+112}$ \\ 
       Counter Arc & 8.77 & 12.06 & $406_{-89}^{+112}$ & $396_{-88}^{+112}$ \\ 
        \hline\hline
    \end{tabular}
    \caption{The mean radii $\langle R \rangle$ of the outflow in the different regions (column 1) of the galaxy based on the surface brightness radial profiles for {\feiiems} (column 2) and {\mgii} (column 3) emission from \citetalias{Shaban202230kpc}.  Columns (4) and (5) are the measured escape velocity at the radii from columns (2) and (3), respectively.}
    \label{tab:mean_radius_outflow}
\end{table}

\begin{landscape}
\setlength{\tabcolsep}{10pt} 
\renewcommand{\arraystretch}{1.5}
\begin{table}
    \centering
    \begin{tabular}{ccccccccccccc}
        \hline\hline
        Region & $\mathrm{W_{out}}$ & $\mathrm{W_{out}}$ & $\mathrm{V_{out}}$ & $\mathrm{V_{out}}$ & $\mathrm{C_{f, out}}$ & $\mathrm{C_{f, out}}$ & $\mathrm{Log_{10}(N_{out})}$ &  $\mathrm{Log_{10}(N_{out})}$ & $\mathrm{V_{90}}$ & $\mathrm{V_{95}}$ & $\mathrm{V_{90}}$ & $\mathrm{V_{95}}$  \\ 
        Name & {\mgii} 2796 & {\feii} 2382 &  {\mgii} 2796 & {\feii} 2382 & {\mgii} & {\feii} & {\mgii} & {\feii} & {\feii} 2382 & {\feii} 2382 &  {\mgii} 2796 & {\mgii} 2796   \\
         & [\AA] & [\AA] & [$\mathrm{km\ s^{-1}}$] & [$\mathrm{km\ s^{-1}}$] &  & & [$\mathrm{cm^{-2}}$] &  [$\mathrm{cm^{-2}}$] & [$\mathrm{km\ s^{-1}}$] & [$\mathrm{km\ s^{-1}}$] & [$\mathrm{km\ s^{-1}}$] & [$\mathrm{km\ s^{-1}}$] \\ 
         (1) & (2) & (3) & (4) & (5) & (6) & (7) & (8) & (9) & (10) & (11) & (12) & (13) \\
        \hline
        \hline
        B1 & $2.40_{-0.12}^{+0.12}$ & $1.63_{-0.05}^{+0.05}$ &     $-193_{-7}^{+7}$ & $-151_{-5}^{+4}$ &     $0.77_{-0.05}^{+0.05}$ & $0.76_{-0.03}^{+0.03}$ &     $14.88_{-0.16}^{+0.23}$ & $15.41_{-0.12}^{+0.11}$ &     $262_{-8}^{+8}$ & $288_{-9}^{+10}$ &     $337_{-14}^{+15}$ & $376_{-18}^{+18}$ \\ 
        E1,1 & $2.61_{-0.06}^{+0.07}$ & $1.95_{-0.03}^{+0.03}$ &     $-178_{-3}^{+3}$ & $-163_{-2}^{+2}$ &     $0.92_{-0.02}^{+0.02}$ & $0.86_{-0.02}^{+0.02}$ &     $15.39_{-0.43}^{+0.37}$ & $15.47_{-0.08}^{+0.10}$ &     $279_{-4}^{+4}$ & $305_{-5}^{+5}$ &     $301_{-5}^{+6}$ & $326_{-7}^{+9}$ \\ 
        E1,2 & $3.74_{-0.24}^{+0.23}$ & $2.61_{-0.09}^{+0.09}$ &     $-251_{-12}^{+11}$ & $-242_{-8}^{+7}$ &     $0.84_{-0.07}^{+0.09}$ & $0.63_{-0.02}^{+0.02}$ &     $14.65_{-0.18}^{+0.41}$ & $15.68_{-0.11}^{+0.10}$ &     $462_{-14}^{+14}$ & $509_{-16}^{+16}$ &     $472_{-27}^{+32}$ & $528_{-39}^{+44}$ \\ 
        U1 & $3.81_{-0.09}^{+0.09}$ & $2.01_{-0.05}^{+0.05}$ &     $-287_{-7}^{+7}$ & $-207_{-5}^{+5}$ &     $0.87_{-0.03}^{+0.04}$ & $0.66_{-0.02}^{+0.02}$ &     $14.75_{-0.06}^{+0.06}$ & $15.35_{-0.04}^{+0.04}$ &     $379_{-10}^{+10}$ & $425_{-11}^{+12}$ &     $531_{-14}^{+15}$ & $608_{-17}^{+18}$ \\ 
        \hline
        U2,1 & $3.27_{-0.10}^{+0.10}$ & $1.94_{-0.05}^{+0.05}$ &     $-244_{-10}^{+10}$ & $-175_{-5}^{+5}$ &     $0.86_{-0.06}^{+0.07}$ & $0.66_{-0.02}^{+0.02}$ &     $14.58_{-0.08}^{+0.08}$ & $15.30_{-0.06}^{+0.05}$ &     $342_{-9}^{+9}$ & $382_{-11}^{+11}$ &     $479_{-19}^{+19}$ & $553_{-23}^{+23}$ \\ 
        U2,2 & $2.59_{-0.12}^{+0.12}$ & $1.91_{-0.05}^{+0.05}$ &     $-206_{-12}^{+10}$ & $-186_{-5}^{+5}$ &     $0.83_{-0.05}^{+0.05}$ & $0.71_{-0.02}^{+0.02}$ &     $14.70_{-0.11}^{+0.12}$ & $15.36_{-0.06}^{+0.06}$ &     $333_{-9}^{+9}$ & $370_{-11}^{+11}$ &     $366_{-21}^{+25}$ & $415_{-27}^{+30}$ \\ 
        U2,3 & $3.49_{-0.17}^{+0.16}$ & $2.34_{-0.09}^{+0.09}$ &     $-191_{-10}^{+9}$ & $-216_{-8}^{+7}$ &     $0.96_{-0.05}^{+0.03}$ & $0.77_{-0.03}^{+0.03}$ &     $14.72_{-0.13}^{+0.15}$ & $15.54_{-0.08}^{+0.09}$ &     $377_{-14}^{+16}$ & $418_{-16}^{+18}$ &     $373_{-19}^{+20}$ & $419_{-24}^{+25}$ \\  
        E2,1 & $3.81_{-0.13}^{+0.12}$ & $2.93_{-0.06}^{+0.06}$ &     $-218_{-6}^{+6}$ & $-194_{-4}^{+4}$ &     $0.95_{-0.04}^{+0.03}$ & $0.85_{-0.02}^{+0.02}$ &     $14.70_{-0.13}^{+0.14}$ & $15.54_{-0.08}^{+0.06}$ &     $380_{-7}^{+7}$ & $421_{-8}^{+8}$ &     $421_{-13}^{+14}$ & $475_{-17}^{+18}$ \\ 
        E2,2 & $3.31_{-0.08}^{+0.08}$ & $2.56_{-0.04}^{+0.04}$ &     $-200_{-4}^{+4}$ & $-165_{-2}^{+2}$ &     $0.95_{-0.02}^{+0.02}$ & $0.95_{-0.01}^{+0.01}$ &     $15.01_{-0.27}^{+0.27}$ & $15.58_{-0.06}^{+0.07}$ &     $305_{-4}^{+4}$ & $332_{-5}^{+5}$ &     $358_{-9}^{+9}$ & $395_{-12}^{+12}$ \\ 
        B2,1 & $1.80_{-0.17}^{+0.15}$ & $1.44_{-0.10}^{+0.11}$ &     $-110_{-9}^{+8}$ & $-120_{-8}^{+7}$ &     $0.88_{-0.11}^{+0.08}$ & $0.72_{-0.07}^{+0.08}$ &     $14.70_{-0.32}^{+0.36}$ & $15.08_{-0.15}^{+0.14}$ &     $230_{-14}^{+15}$ & $255_{-15}^{+17}$ &     $208_{-13}^{+17}$ & $231_{-14}^{+19}$ \\ 
        B2,2 & $2.79_{-0.14}^{+0.14}$ & $1.51_{-0.07}^{+0.07}$ &     $-187_{-7}^{+7}$ & $-150_{-6}^{+5}$ &     $0.90_{-0.05}^{+0.05}$ & $0.67_{-0.04}^{+0.04}$ &     $14.84_{-0.19}^{+0.25}$ & $15.33_{-0.13}^{+0.15}$ &     $266_{-10}^{+11}$ & $290_{-11}^{+12}$ &     $332_{-15}^{+15}$ & $369_{-19}^{+20}$ \\ 
        \hline
        Image 3,1 & $1.88_{-0.13}^{+0.14}$ & $1.69_{-0.04}^{+0.04}$ &     $-206_{-8}^{+8}$ & $-161_{-3}^{+3}$ &     $0.53_{-0.05}^{+0.07}$ & $0.67_{-0.02}^{+0.02}$ &     $14.93_{-0.34}^{+0.53}$ & $15.40_{-0.07}^{+0.06}$ &     $297_{-6}^{+6}$ & $329_{-7}^{+7}$ &     $368_{-14}^{+15}$ & $406_{-18}^{+19}$ \\ 
        Image 3,2 & $2.14_{-0.14}^{+0.24}$ & $1.72_{-0.03}^{+0.03}$ &     $-259_{-8}^{+7}$ & $-240_{-4}^{+4}$ &     $0.42_{-0.03}^{+0.06}$ & $0.50_{-0.01}^{+0.01}$ &     $15.44_{-0.24}^{+0.27}$ & $15.69_{-0.04}^{+0.04}$ &     $424_{-8}^{+8}$ & $469_{-9}^{+9}$ &     $489_{-14}^{+15}$ & $536_{-18}^{+18}$ \\ 
        Image 3,3 & $2.52_{-0.13}^{+0.13}$ & $1.94_{-0.06}^{+0.06}$ &     $-239_{-8}^{+8}$ & $-218_{-6}^{+6}$ &     $0.93_{-0.08}^{+0.05}$ & $0.58_{-0.02}^{+0.02}$ &     $14.25_{-0.08}^{+0.12}$ & $15.66_{-0.11}^{+0.12}$ &     $392_{-11}^{+12}$ & $427_{-14}^{+14}$ &     $371_{-10}^{+10}$ & $399_{-10}^{+12}$ \\ 
        \hline
        Counter arc & $4.03_{-0.08}^{+0.08}$ & $2.79_{-0.04}^{+0.04}$ &     $-257_{-6}^{+5}$ & $-255_{-3}^{+3}$ &     $0.88_{-0.02}^{+0.02}$ & $0.77_{-0.01}^{+0.01}$ &     $14.87_{-0.04}^{+0.05}$ & $15.71_{-0.02}^{+0.02}$ &     $448_{-6}^{+6}$ & $496_{-7}^{+7}$ &     $496_{-11}^{+11}$ & $563_{-13}^{+14}$ \\
        \hline
        \hline
    \end{tabular}
    \caption{Measured outflow properties of the absorption lines from the \texttt{emcee} model realizations for both {\mgii} and {\feii}. Column (1) is the name of the pseudo-slits and their corresponding image (region+image). Columns (2) and (3) are the measured rest-frame equivalent widths for the outflow component for {\mgii} 2796 and {\feii} 2382, respectively. Columns (4) and (5) are the measured median outflow velocity for {\mgii} 2796 and {\feii} 2382, respectively. Columns (6) and (7) are the measured outflow covering fraction for {\mgii} 2796 and {\feii} 2382, respectively. Columns (8) and (9) are the measured column densities for the outflow component for {\mgii} and {\feii}, respectively. Columns (10) and (11) are the 90$\rm{^th}$ 95$\rm{^th}$ percentile velocities for {\feii} 2382, respectively. Columns (12) and (13) are the 90$\rm{^th}$ 95$\rm{^th}$ percentile velocities for {\mgii} 2796, respectively. The horizontal lines in the Table represent the transitions from one image to another in the image-plane.}
    \label{table:outflow_properties}
\end{table}
\end{landscape}
\setlength{\tabcolsep}{10pt} 
\renewcommand{\arraystretch}{1.5}
\begin{table*}
    \centering
    \begin{tabular}{ccccccccc}
        \hline\hline

        Region & $\dot{M}_{out}$ & $\dot{M}_{out}$ & $Log_{10}(\dot{p}_{out} )$ & $Log_{10}(\dot{p}_{out})$ & $Log_{10}(\dot{E}_{out} )$ & $Log_{10}(\dot{E}_{out})$ & $\dot{M}_{esc}$ & $\dot{M}_{esc}$ \\ 
        Name & {\mgii} & {\feii} & {\mgii} & {\feii} & {\mgii} & {\feii} & {\mgii} & {\feii} \\
         & [$M_{\odot} yr^{-1}$] & [$M_{\odot} yr^{-1}$] & [dynes] & [dynes] &  [$\mathrm{erg\ s^{-1}}$] & [$\mathrm{erg\ s^{-1}}$] & [$M_{\odot} yr^{-1}$] & [$M_{\odot} yr^{-1}$]  \\
         (1) & (2) & (3) & (4) & (5) & (6) & (7) & (8) & (9)\\
        \hline\hline
        B1 & $72_{-23}^{+50}$ & $101_{-24}^{+29}$ & $34.94_{-0.17}^{+0.23}$ & $34.98_{-0.12}^{+0.11}$ & $41.93_{-0.18}^{+0.23}$ & $41.86_{-0.13}^{+0.12}$ & $0_{-0}^{+2}$ & $0_{-0}^{+1}$ \\ 
E1,1 & $196_{-122}^{+263}$ & $103_{-18}^{+26}$ & $35.34_{-0.42}^{+0.37}$ & $35.02_{-0.09}^{+0.10}$ & $42.29_{-0.42}^{+0.37}$ & $41.93_{-0.09}^{+0.10}$ & $0_{-0}^{+1}$ & $0_{-0}^{+1}$ \\ 
E1,2 & $51_{-18}^{+77}$ & $254_{-60}^{+65}$ & $34.91_{-0.19}^{+0.39}$ & $35.59_{-0.12}^{+0.10}$ & $42.01_{-0.20}^{+0.39}$ & $42.67_{-0.12}^{+0.11}$ & $3_{-1}^{+7}$ & $8_{-4}^{+11}$ \\ 
U1 & $63_{-8}^{+9}$ & $102_{-9}^{+11}$ & $35.05_{-0.06}^{+0.06}$ & $35.12_{-0.05}^{+0.05}$ & $42.21_{-0.07}^{+0.07}$ & $42.14_{-0.05}^{+0.05}$ & $4_{-2}^{+4}$ & $3_{-2}^{+7}$ \\ 
\hline
U2,1 & $36_{-6}^{+8}$ & $77_{-10}^{+10}$ & $34.74_{-0.09}^{+0.09}$ & $34.92_{-0.06}^{+0.06}$ & $41.82_{-0.10}^{+0.11}$ & $41.87_{-0.07}^{+0.06}$ & $3_{-2}^{+3}$ & $0_{-0}^{+3}$ \\ 
U2,2 & $40_{-9}^{+12}$ & $95_{-13}^{+14}$ & $34.72_{-0.11}^{+0.11}$ & $35.04_{-0.07}^{+0.06}$ & $41.73_{-0.12}^{+0.12}$ & $42.01_{-0.07}^{+0.07}$ & $0_{-0}^{+2}$ & $1_{-1}^{+5}$ \\ 
U2,3 & $39_{-10}^{+17}$ & $164_{-29}^{+35}$ & $34.67_{-0.14}^{+0.16}$ & $35.35_{-0.09}^{+0.09}$ & $41.65_{-0.15}^{+0.17}$ & $42.38_{-0.09}^{+0.09}$ & $1_{-0}^{+4}$ & $2_{-2}^{+10}$ \\ 
E2,1 & $49_{-13}^{+18}$ & $147_{-24}^{+23}$ & $34.82_{-0.14}^{+0.14}$ & $35.25_{-0.08}^{+0.06}$ & $41.86_{-0.15}^{+0.14}$ & $42.24_{-0.08}^{+0.07}$ & $2_{-1}^{+7}$ & $2_{-1}^{+10}$ \\ 
E2,2 & $91_{-43}^{+77}$ & $137_{-17}^{+26}$ & $35.06_{-0.27}^{+0.26}$ & $35.15_{-0.06}^{+0.08}$ & $42.06_{-0.28}^{+0.26}$ & $42.07_{-0.06}^{+0.08}$ & $1_{-1}^{+4}$ & $0_{-0}^{+8}$ \\ 
B2,1 & $27_{-14}^{+38}$ & $37_{-11}^{+15}$ & $34.27_{-0.32}^{+0.39}$ & $34.45_{-0.16}^{+0.16}$ & $41.01_{-0.32}^{+0.41}$ & $41.22_{-0.17}^{+0.17}$ & $0_{-0}^{+0}$ & $0_{-0}^{+0}$ \\ 
B2,2 & $65_{-23}^{+49}$ & $85_{-22}^{+34}$ & $34.88_{-0.19}^{+0.24}$ & $34.90_{-0.14}^{+0.15}$ & $41.85_{-0.20}^{+0.24}$ & $41.78_{-0.14}^{+0.15}$ & $0_{-0}^{+2}$ & $0_{-0}^{+1}$ \\
\hline
Image 3,1 & $137_{-76}^{+331}$ & $183_{-28}^{+29}$ & $35.25_{-0.36}^{+0.54}$ & $35.27_{-0.07}^{+0.07}$ & $42.26_{-0.37}^{+0.54}$ & $42.17_{-0.08}^{+0.07}$ & $1_{-1}^{+3}$ & $0_{-0}^{+6}$ \\ 
Image 3,2 & $554_{-234}^{+468}$ & $527_{-47}^{+48}$ & $35.95_{-0.24}^{+0.27}$ & $35.90_{-0.04}^{+0.04}$ & $43.07_{-0.24}^{+0.27}$ & $42.98_{-0.04}^{+0.04}$ & $5_{-3}^{+4}$ & $8_{-4}^{+14}$ \\ 
Image 3,3 & $33_{-5}^{+10}$ & $445_{-104}^{+151}$ & $34.70_{-0.08}^{+0.11}$ & $35.79_{-0.12}^{+0.13}$ & $41.78_{-0.08}^{+0.11}$ & $42.82_{-0.13}^{+0.14}$ & $1_{-1}^{+10}$ & $6_{-5}^{+19}$ \\
\hline
Counter arc & $169_{-17}^{+20}$ & $527_{-29}^{+30}$ & $35.44_{-0.05}^{+0.05}$ & $35.93_{-0.03}^{+0.03}$ & $42.54_{-0.06}^{+0.06}$ & $43.03_{-0.03}^{+0.03}$ & $38_{-19}^{+18}$ & $17_{-9}^{+26}$ \\
        \hline\hline
    \end{tabular}
    \caption{Measured mass outflow rates $\dot{M}_{out}$, momentum outflow flux $\dot{p}_{out}$, and energy outflow flux $\dot{E}_{out}$ for \rcsa. Column (1) is the name of the pseudo-slits and their corresponding image (region+image). Columns (2) and (3) represent the measured mass outflow rates from {\mgii} and {\feii}, respectively. Columns (4) and (5) are the outflow momentum flux for {\mgii} and {\feii}, respectively. Columns (6) and (7) are the outflow energy flux for {\mgii} and {\feii}, respectively. Columns (8) and (9) represent the escaping mass outflow rate for {\mgii} and {\feii}, respectively. The horizontal lines in the Table represent the transitions from one image to another in the image-plane.}
    \label{table:M_P_E_dot}
\end{table*}

\begin{landscape}
\setlength{\tabcolsep}{10pt}
\renewcommand{\arraystretch}{1.5}
\begin{table}
    \centering
    \begin{tabular}{cccccccccccc}
        \hline\hline 
        Region & $\mathrm{W_{ems,1}}$ & $\mathrm{W_{ems,2}}$ & $\mathrm{W_{ems,1}}$ & $\mathrm{W_{ems,2}}$ & $\mathrm{W_{ems}}$ & $\mathrm{V_{ems,1}}$ & $\mathrm{V_{ems,2}}$ & $\mathrm{V_{ems}}$ & $\mathrm{b_{D,ems1}}$ &  $\mathrm{b_{D,ems2}}$ & $\mathrm{b_{D,ems}}$\\ 
        Name & {\mgii} 2796 & {\mgii} 2796 & {\mgii} 2803 & {\mgii} 2803 & {\feii} 2626 & {\mgii} & {\mgii} & {\feii} & {\mgii} & {\mgii} & {\feii} \\
         & [{\AA}] & [{\AA}] & [{\AA}] & [{\AA}] & [{\AA}] & [$\mathrm{km\ s^{-1}}$] & [$\mathrm{km\ s^{-1}}$] & [$\mathrm{km\ s^{-1}}$]& [$\mathrm{km\ s^{-1}}$]& [$\mathrm{km\ s^{-1}}$]& [$\mathrm{km\ s^{-1}}$] \\
        (1) & (2) & (3) & (4) & (5) & (6) & (7) & (8) & (9) & (10) & (11) & (12)\\
        \hline
        B1 & $-1.95_{-0.67}^{+0.38}$ & $-1.13_{-0.33}^{+0.23}$ & $-1.39_{-0.31}^{+0.27}$ & $-1.26_{-0.27}^{+0.25}$ & $-0.64_{-0.07}^{+0.07}$ & $138_{-14}^{+9}$ & $422_{-49}^{+47}$ & $-13_{-8}^{+8}$ & $72_{-12}^{+16}$ & $264_{-63}^{+74}$ & $105_{-4}^{+8}$ \\ 
        E1,1 & $-4.05_{-1.14}^{+0.69}$ & $-2.23_{-0.63}^{+0.42}$ & $-1.03_{-0.18}^{+0.15}$ & $-0.90_{-0.13}^{+0.11}$ & $-1.39_{-0.06}^{+0.06}$ & $107_{-33}^{+23}$ & $592_{-16}^{+16}$ & $-16_{-4}^{+4}$ & $161_{-21}^{+24}$ & $135_{-22}^{+29}$ & $139_{-7}^{+7}$ \\ 
        E1,2 & $-4.42_{-1.62}^{+0.93}$ & $-3.23_{-1.03}^{+0.56}$ & $-1.00_{-0.69}^{+0.50}$ & $-0.80_{-0.49}^{+0.41}$ & $-3.04_{-0.14}^{+0.14}$ & $148_{-21}^{+9}$ & $382_{-39}^{+88}$ & $8_{-4}^{+5}$ & $96_{-22}^{+31}$ & $167_{-61}^{+83}$ & $137_{-8}^{+8}$ \\ 
        U1 & $-4.62_{-0.29}^{+0.26}$ & $-3.22_{-0.24}^{+0.22}$ & $-0.63_{-0.12}^{+0.10}$ & $-0.56_{-0.09}^{+0.08}$ & $-1.45_{-0.06}^{+0.06}$ & $78_{-6}^{+5}$ & $533_{-18}^{+15}$ & $-10_{-3}^{+3}$ & $99_{-6}^{+6}$ & $86_{-20}^{+27}$ & $112_{-6}^{+6}$ \\ 
        \hline
        U2,1 & $-7.14_{-0.95}^{+0.88}$ & $-4.14_{-0.48}^{+0.45}$ & $-0.93_{-0.22}^{+0.20}$ & $-0.85_{-0.20}^{+0.18}$ & $-1.71_{-0.06}^{+0.06}$ & $90_{-8}^{+8}$ & $372_{-32}^{+48}$ & $25_{-3}^{+3}$ & $88_{-6}^{+6}$ & $259_{-57}^{+76}$ & $110_{-5}^{+5}$ \\ 
        U2,2 & $-6.11_{-1.20}^{+0.91}$ & $-4.48_{-0.79}^{+0.59}$ & $-0.46_{-0.27}^{+0.22}$ & $-0.36_{-0.22}^{+0.17}$ & $-1.90_{-0.07}^{+0.07}$ & $72_{-13}^{+11}$ & $503_{-126}^{+172}$ & $-11_{-3}^{+3}$ & $93_{-10}^{+10}$ & $271_{-157}^{+91}$ & $105_{-3}^{+5}$ \\ 
        U2,3 & $-8.96_{-3.07}^{+1.97}$ & $-6.61_{-2.11}^{+1.35}$ & $-1.35_{-0.62}^{+0.51}$ & $-1.10_{-0.48}^{+0.43}$ & $-2.45_{-0.11}^{+0.11}$ & $59_{-33}^{+23}$ & $425_{-64}^{+125}$ & $-19_{-3}^{+3}$ & $140_{-24}^{+28}$ & $264_{-100}^{+96}$ & $101_{-1}^{+2}$ \\ 
        E2,1 & $-3.59_{-1.66}^{+0.75}$ & $-2.03_{-0.88}^{+0.40}$ & $-1.22_{-0.38}^{+0.28}$ & $-0.92_{-0.19}^{+0.17}$ & $-1.75_{-0.10}^{+0.10}$ & $148_{-48}^{+23}$ & $680_{-26}^{+24}$ & $-1_{-5}^{+5}$ & $141_{-27}^{+41}$ & $122_{-27}^{+34}$ & $142_{-10}^{+10}$ \\ 
        E2,2 & $-4.86_{-0.77}^{+0.92}$ & $-2.98_{-0.52}^{+0.60}$ & $-0.93_{-0.25}^{+0.19}$ & $-0.75_{-0.16}^{+0.12}$ & $-1.77_{-0.09}^{+0.09}$ & $25_{-17}^{+24}$ & $534_{-22}^{+18}$ & $-66_{-6}^{+6}$ & $152_{-19}^{+15}$ & $115_{-32}^{+50}$ & $174_{-11}^{+11}$ \\ 
        B2,1 & $-5.68_{-1.10}^{+0.67}$ & $-2.94_{-0.56}^{+0.34}$ & $-0.20_{-0.27}^{+0.15}$ & $-0.15_{-0.21}^{+0.11}$ & $-1.32_{-0.15}^{+0.14}$ & $165_{-15}^{+9}$ & $644_{-182}^{+94}$ & $19_{-8}^{+8}$ & $93_{-12}^{+14}$ & $100_{-41}^{+151}$ & $122_{-13}^{+16}$ \\ 
        B2,2 & $-2.73_{-1.01}^{+0.36}$ & $-1.46_{-0.51}^{+0.20}$ & $-1.15_{-0.36}^{+0.29}$ & $-0.91_{-0.22}^{+0.21}$ & $-1.41_{-0.11}^{+0.11}$ & $143_{-31}^{+12}$ & $679_{-41}^{+42}$ & $-15_{-7}^{+7}$ & $99_{-16}^{+28}$ & $124_{-42}^{+41}$ & $107_{-5}^{+9}$ \\ 
        \hline
        Image 3,1 & $-3.26_{-0.67}^{+0.66}$ & $-1.76_{-0.29}^{+0.26}$ & $-0.52_{-0.12}^{+0.09}$ & $-0.46_{-0.10}^{+0.08}$ & $-0.86_{-0.06}^{+0.06}$ & $111_{-8}^{+7}$ & $543_{-19}^{+15}$ & $-44_{-5}^{+5}$ & $93_{-6}^{+6}$ & $80_{-21}^{+45}$ & $135_{-9}^{+9}$ \\ 
        Image 3,2 & $-2.97_{-0.61}^{+0.78}$ & $-1.70_{-0.22}^{+0.21}$ & $-1.39_{-0.17}^{+0.22}$ & $-1.37_{-0.17}^{+0.22}$ & $-0.89_{-0.04}^{+0.04}$ & $109_{-5}^{+4}$ & $383_{-24}^{+41}$ & $-3_{-4}^{+4}$ & $88_{-6}^{+9}$ & $228_{-24}^{+21}$ & $128_{-7}^{+7}$ \\ 
        Image 3,3 & $-2.28_{-0.38}^{+0.28}$ & $-1.21_{-0.19}^{+0.15}$ & $-1.06_{-0.28}^{+0.23}$ & $-0.83_{-0.19}^{+0.16}$ & $-0.98_{-0.09}^{+0.09}$ & $15_{-7}^{+6}$ & $422_{-21}^{+18}$ & $-73_{-8}^{+8}$ & $68_{-11}^{+14}$ & $88_{-26}^{+43}$ & $105_{-3}^{+6}$ \\ 
        \hline
        Counter arc & $-3.06_{-0.28}^{+0.24}$ & $-1.99_{-0.23}^{+0.19}$ & $-1.35_{-0.25}^{+0.24}$ & $-1.28_{-0.23}^{+0.23}$ & $-1.45_{-0.06}^{+0.05}$ & $87_{-7}^{+6}$ & $385_{-36}^{+46}$ & $-59_{-3}^{+3}$ & $113_{-9}^{+9}$ & $296_{-56}^{+60}$ & $130_{-6}^{+6}$ \\ 
        \hline
        \hline
    \end{tabular}
    \caption{Outflow emission lines properties. Column (1) is the name of the pseudo-slits and their corresponding image (region+image). Columns (2) and (3) represent the measured rest-frame equivalent width for the primary and secondary emission components for {\mgii} 2796, respectively. Columns (4) and (5) are the rest-frame equivalent width measurements for the primary and secondary emission components for {\mgii} 2803, respectively. Column (6) is the rest-frame equivalent width measurement for {\feiiems} 2626. Columns (7), (8) and (9) represent the measured emission velocity for the primary and secondary component for {\mgii}, and {\feiiems} emission, respectively. Columns (10), (11), and (12) are the measured Doppler parameter for the primary and secondary {\mgii}  emission components, and the {\feiiems} emission. The horizontal lines in the Table represent the transitions from one image to another in the image-plane.}
    \label{tab:outflows_emission}
\end{table}
\end{landscape}


\bsp	
\label{lastpage}
\end{document}